\documentclass{aa}
\usepackage{graphicx}
\usepackage{txfonts}
 \usepackage{graphicx}
 \usepackage[normalem]{ulem}
 \useunder{\uline}{\ul}{}
 \usepackage{lscape}
 \usepackage{natbib}
 \usepackage{xcolor}
 \usepackage[switch]{lineno}
 
\begin{document} 

\title{Revisiting empirical solar energetic particle scaling relations}

   \subtitle{I. Solar flares }

   \author{Athanasios Papaioannou
          \inst{1}
          \and
          Konstantin Herbst\inst{2}
        \and
          Tobias Ramm \inst{3}
        \and
          Edward W. Cliver \inst{4}
        \and
          David Lario \inst{5}
       \and
          Astrid M. Veronig \inst{6}
   }       

   \institute{Institute for Astronomy, Astrophysics, Space Applications and Remote Sensing (IAASARS), National Observatory of Athens, I. Metaxa \& Vas. Pavlou St., 15236 Penteli, Greece\\
              \email{atpapaio@astro.noa.gr}
         \and
             Institut f\"ur Experimentelle und Angewandte Physik, Christian-Albrechts-Universit\"at zu Kiel, 24118 Kiel, Germany
             \and
        Institut f\"ur Theoretische Physik und Astrophysik, Christian-Albrechts-Universit\"at zu Kiel, 24118 Kiel, Germany
       \and
        National Solar Observatory, 3665 Discovery Drive, Boulder, CO 80303, USA
       \and
        NASA, Goddard Space Flight Center, Heliophysics Science Division, Greenbelt, MD 20771, USA
        \and
       Institute of Physics \& Kanzelh\"{o}he Observatory for Solar and Environmental Research, University of Graz, A-8010 Graz, Austria}

  \titlerunning{Solar Scaling Relations: Part I}  

 
  \abstract
   {}
  {The possible influence of solar superflares on the near-Earth space radiation environment are assessed through the investigation
of scaling laws between the peak proton flux and fluence of solar energetic particle (SEP) events with the solar flare soft X-ray peak
photon flux.}
   {We compiled a catalog of 65 well-connected (W20-90) SEP events during the last three solar cycles covering a period of $\sim$34 years (1984--2020) that were associated with flares of class $\geq$C6.0, and investigated the statistical relations between the recorded peak proton fluxes ($I_{P}$) and the fluences ($F_{P}$) at a set of integral energies from E $>$10, $>$30, and $>$60 to $>$100 MeV versus the associated solar flare peak soft X-ray flux in the 1–8 Å band ($F_{SXR}$). Based on the inferred relations, we calculated the integrated energy dependence of the peak proton flux ($I_{P}$) and fluence ($F_{P}$) of the SEP events, assuming that they follow an inverse power law with respect to energy. Finally, we made use of simple physical assumptions, combining our derived scaling laws, and estimated the upper limits for $I_{P}$ and $F_{P}$ focusing on the flare associated with the strongest ground level
enhancement (GLE)  directly observed to date (GLE 05 on 23 February 1956), and that inferred for the cosmogenic radionuclide-based SEP event of AD774/775.}
   {A scaling law relating $I_{P}$ and $F_{P}$ to the solar soft X-ray peak intensity ($F_{SXR}$) as $\propto$~$F_{SXR}^{5/6}$ for a flare with a $F_{SXR}$ = X600 (in the revised scale) is consistent with values of $F_{P}$ inferred for the cosmogenic nuclide event of AD774/775.  }
   {}


   \maketitle


\section{Introduction}
\label{sec:intro}
The radiation environment in a planetary star system is driven by its host star \citep{2003ApJ...598L.121L, 2020IJAsB..19..136A}. In the case of the Solar System, the Sun determines this radiation environment  \citep{2021LRSP...18....4T} since it is the source of solar energetic particles (SEPs), while also modulating the incoming galactic cosmic ray (GCR) flux. SEP protons are accelerated at both solar flares and coronal mass ejections (CMEs) \citep{2010JGRA..115.8101C,2016JSWSC...6A..42P,2021LNP...978.....R}. SEP events that are limited in duration reach small peak intensities and have narrow emission cones; they are thought to be associated with solar flares and type III radio bursts \citep[see, e.g.,][]{2021LNP...978.....R}. On the other hand, high-energy SEP events, which can last for several days, achieve significant peak fluxes, have a broad cone of emission, and are thought to be associated with CMEs and type II radio bursts \citep[e.g.,][]{desai2016large}.  The fact that high-energy protons can be accelerated both during the impulsive phase of flares and at CME-driven shocks   \citep[see, e.g.,][]{forrest1985neutral, 1987ApJ...318..913C, 2010JGRA..115.8101C,2016JSWSC...6A..42P} complicates the interpretation of the mechanisms responsible for the acceleration, injection, and propagation of SEPs in the interplanetary (IP) medium \citep{klein2017acceleration}, although the preponderant evidence favors CME-driven shocks as the dominant source of high-energy protons in the most intense large SEP events \citep[e.g.,][]{desai2016large, cliver2022}.

Large SEP events measured near Earth have been recorded by spacecraft over the last 60 years (for an example of the last three solar cycles, see, e.g., Fig. 15 (A) in \citealt{2016JSWSC...6A..42P}). Singular intense events, particularly at low ( $<$30 MeV) energies,  termed ``rogue'' SEP events by \cite{kallenrode2001rogue}, occurred on 14 July 1959 \citep{bazilevskaya2010solar}, 4 August 1972 \citep{2013AIPC.1539..215L,knipp2018little}, 19 October 1989 \citep{vainio2003generation,2013AIPC.1539..215L}, and 14 July 2000 \citep{belov2001bastille, 2013AIPC.1539..215L, mishev2016analysis}. Such events are associated with multiple CMEs and converging shocks. In particular, the most intense SEP event identified so far during the modern space era occurred on 4 August 1972 with a peak proton flux at E$>$ 10 MeV reaching 6 $\times$ 10$^{4}$ pfu \citep{2004AnGeo..22.2255K}. The omnidirectional integrated fluence of this compound event at an integral energy of E$>$ 30 MeV was estimated to be 5 $\times$ 10$^{9}$ cm$^{ -2}$ \citep{smart2006carrington} and subsequently 8.4 $\times$ 10$^{9}$ cm$^{ -2}$ \citep{2014JSWSC...4A..20J}. A fraction of these large generally soft-spectrum \citep{Cliver_etal_2020} SEP events, as well as numerous other events with harder spectra, can reach such high energies that particles can interact with Earth's atmosphere, and the subproducts are recorded on the ground as significant enhancements above the background GCR flux by neutron monitors \citep[NMs;][]{2011AdSpR..47.2210M}. These events are termed ground level enhancements (GLEs); they reach very high energies ($\ge$ 1-2 GeV) and pose a serious threat for humans and infrastructure \citep{shea2012space}. Since 1956 a total of 73 GLEs have been reported by the global NM network\footnote{\url{https://gle.oulu.fi/}} \citep{2017SoPh..292..176P,anastasiadis2019solar,2022A&A...660L...5P}. Investigating the historical records of solar and geospace observations, researchers attempted to quantify one of the most extreme events that has ever been released by our Sun, known as the Carrington event, that occurred on 1–2 September 1859 \citep{2013JSWSC...3A..31C}. These authors estimated an omnidirectional fluence for the integral energy of E$>$ 30 MeV of $\sim$ 1.1 $\times$ 10$^{10}$ protons cm$^{-2}$, which exceeds the relevant estimates of fluence of the modern era rogue events by a factor of $\sim$1.4. The Carrington event and its corresponding particle fluence were seen as the worst-case estimate of radiation hazard in the near-Earth environment that the Sun is capable of producing \citep{miroshnichenko2014extreme}. However, with the help of cosmogenic radionuclide records, it became clear that much more extreme events (e.g., the event around AD774/775) might have occurred on the Sun \citep{miyake2012signature, 2013A&A...552L...3U}\footnote{At present, analyses of ice cores for the $^{36}$Cl cosmogenic nuclide have revealed no evidence for a significant low-energy SEP event in 1859 \citep{cliver2022}}. 

The soft X-ray (SXR) peak flux of solar flares,  regularly monitored since the mid-1970s by the Geostationary Operational Environmental Satellite Program (GOES) in the 1-8 {\AA} (long) passband, has been widely used by the scientific community. Flares that are associated with intense variations in the radiation environment are categorized into three X-ray flare classes, namely C-, M-, and X-class flares\footnote{The  ranges vary between $10^{-6} - 10^{-5}, 10^{-5} - 10^{-4}, $ and  above $10^{-4}$ W/m$^{2}$ for the C-, M-, and X-classes, respectively}. The largest solar flare that has ever been observed on the Sun in the modern era of spacecraft measurements occurred on 4 November 2003 and resulted in the saturation of the GOES X-ray detector. Its magnitude was estimated by linear extrapolation to be $\sim$X35 \citep{2004AAS...204.4713K,2013JSWSC...3A..31C} and, more recently, X30 (Hudson et al. 2022 in preparation). Between 1976 and 2020 there have been 22 solar flares with a magnitude $\ge$ X10 \citep[see Table 4 in][]{2020ApJ...903...41C}. Research focusing on the largest SXR flares on the Sun provides estimates that reach up to several times X100. For example, \cite{2018ApJ...853...41T} indicated that for the largest active regions (ARs), flares with $\sim$X500 magnitude could be produced. Recently, from consideration of various worst-case estimates of the most intense solar flare based on the largest spot group observed in the last $\sim$150 years (6132 millionths of a solar hemisphere on 8 April, 1947; \citealt{cliver2022}) obtained a consensus value of $\sim$X200 (with bolometric energy $\sim$1.5 $\times$ 10$^{33}$ erg). 

Several statistical studies point to an empirical relation between the SXR flare peak flux and the achieved peak proton flux and fluence of the resulting SEP events \citep[see, e.g.,][]{1982JGR....87.3439K, 2001JGR...10620947K, belov2005proton, 2010JGRA..115.8101C, 2016JSWSC...6A..42P}. Recent studies that investigated such correlations for a set of integral proton energies showed that the correlation of the SEP peak proton flux or the SEP fluence with the flare SXR peak flux was reasonably stable (correlation coefficient $\sim$ 0.43) for all such energies considered \citep[see, e.g.,][]{2015SoPh..290..841D,2016JSWSC...6A..42P}.

In this work we analyze the statistical relations among the SEP peak proton fluxes and omnidirectional fluences of a well-defined catalog of (initially) 67 events measured at 1 AU by GOES between 1984 and 2017 for a set of integral energies spanning E$>$ 10, E$>$ 30, E$>$ 60, and E$>$100 MeV and the SXR peak fluxes of their parent solar events. \citet{2016ApJ...833L...8T} deduced that the upper limit for the peak proton flux ($I_{P}$) of E$>$10 MeV is proportional to the SXR flux ($I_{P} \propto ~F_{SXR}^{5/6}$). Based upon this result and expanding their argumentation, we derive upper limits and scaling relations among the SEP peak flux ($I_{P}$) at each integral energy (from E$>$10 to E$>$100 MeV) and the SXR peak flux ($F_{SXR}$). We extend these relations to also incorporate  the fluence of the SEP events ($F_{P}$). We further calculate the integral SEP peak flux and fluence spectra for the events in our sample, which are assumed to follow an inverse power law. We compare our findings with the most extreme peak proton fluxes and fluences that have ever been recorded for a GLE in modern times \citep{koldobskiy2021new}, namely the strong hard-spectrum GLE that occurred on 23 February 1956 (GLE05), and with the superflare of AD774/775.

Based on estimates of the largest possible SXR flare, the obtained scaling laws, and the observations used in this work, we estimate the most intense SEP proton fluxes and fluences that the Sun can produce, and the corresponding SEP spectra (for both quantities). In the concluding section of this study the implications of the effects of solar superflares on the radiation environment are put forward and discussed.

\section{Data sets}
\label{sec:data}
The SEP data were scanned from 1984 to 2020, aiming at identifying well-connected SEP events (W20-90$^{o}$) that reached integral energies of E$>$100 MeV. No such events occurred between 2018 and 2020. We identified 67 well-connected SEP events between 1984 and 2017 that extended from E$>$ 10 MeV to E$>$ 100 MeV. However, two of these events had to be excluded from our analysis  because our event selection is based on that of \citet{2019A&A...621A..67H}, who considered SEP events whose origin is associated with X-ray solar flares of class $\geq$C6.0. The solar flare characteristics of the remaining 65 events were obtained from the online repository of the National Oceanic and Atmospheric Administration (NOAA).\footnote{\url{https://www.ngdc.noaa.gov/stp/space-weather/solar-data/solar-features/solar-flares/x-rays/goes/}} We note that in order to obtain  accurate SXR fluxes, proper scaling was applied (using a multiplicative factor of 1/0.7 for the GOES 1--8 \AA\ channel).\footnote{\url{https://www.ngdc.noaa.gov/stp/satellite/goes/doc/GOES_XRS_readme.pdf}}  The saturated strong X-class event values also need further attention; their re-scaled SXR classes are taken from Hudson et al. (2022; in preparation).  

For the SEP events between 1986 and 2017 we used the corrected\footnote{\url{https://ngdc.noaa.gov/stp/satellite/goes/datanotes.html}} GOES/EPS data.\footnote{\url{https://satdat.ngdc.noaa.gov/sem/goes/data/avg/}} For each of the identified SEP events we were able to identify the peak intensity (in units of protons cm$^{-2}$ sr$^{-1}$ s$^{-1}$, 5 min averages) in their prompt component, similarly to \cite{2014JGRA..119.4185L}, which is understood as the maximum intensity observed shortly after the onset of the event in situ and several hours or days before the particle enhancement commonly associated with the arrival of an interplanetary shock (if any) at the spacecraft (i.e., energetic storm particles or the ESP component were excluded). For some events, the maximum intensity in the prompt component is observed as a plateau in the time--intensity profile before the local enhancement associated with the passage of shocks \citep{1998ApJ...504.1002R}. In these cases the peak intensity is taken as the maximum value of the intensity plateau. It should also be noted that different GOES satellites may record different time profiles because of their positions in space and calibration dissimilarities. In this work we scanned through all the available recordings and spacecraft, selecting those GOES satellites that had the highest peak flux. Moreover, from the retrieved integral proton intensities, we   computed the omni-directional time-integrated fluences (in units of protons cm$^{-2}$) by integrating each channel throughout the SEP event and multiplying the result by 4$\pi$ \citep{2011SpWea...911003L}. Since the repository mentioned above provides no integral data for three SEP events marked earlier (i.e., in 1984-1985), we used the peak proton flux and fluence values that are included in the paper by \cite{2016JSWSC...6A..42P}. The obtained GOES peak proton fluxes and  fluences, and the particular GOES satellite used for each SEP event are listed in Appendix \ref{appendix:C}.

\section{Scaling relations}
\label{sec:scaling}
\subsection{Soft X-ray flare flux and peak proton fluxes}
\label{sec:1}
As a first step, we obtained the scaling relations between the SXR magnitude of the solar flares ($F_{SXR}$) and the peak proton flux ($I_{P}$) for the integral energies E$>$ 10 MeV, E$>$ 30 MeV, E$>$ 60 MeV, and E$>$ 100 MeV. The proportionality relationships between $F_{SXR}$ and $I_{P}$ have been primarily investigated for an integral energy of E$>$ 10 MeV \citep[see][]{2007SoPh..246..457B, 2012ApJ...756L..29C, 2019A&A...621A..67H}. However, our goal is to further investigate these relations up to higher energies, quantify whether they vary, and thus identify and quantify the conditions that lead to a potential variability.

\begin{figure}[t!]
\centering
\includegraphics[width=\columnwidth]{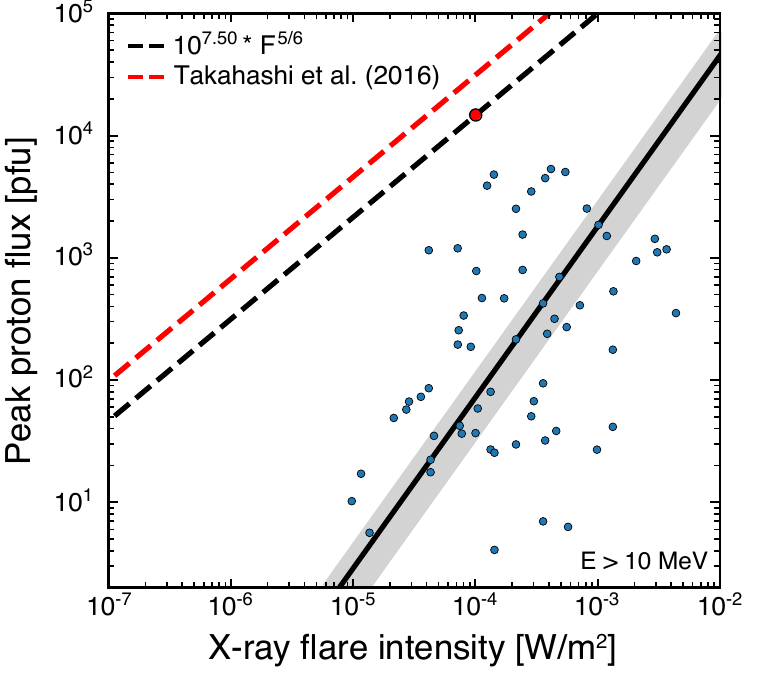}
\caption{Scatter plot of the E$>$10 MeV peak proton flux vs. the X-ray flare intensity of our new sample (blue dots). The solid black line corresponds to the RMA regression fit of our sample corresponding to $I_{P}$ $\propto$ $F_{SXR}^{\beta}$ with $\beta$=1.40$\pm$0.19. The red dashed line gives the upper solar limit from \citet{2016ApJ...833L...8T} based on a scaling law of $I_{P}$ $\propto$ $F_{SXR}^{5/6}$. In the figure, this relation was re-scaled to the upper-point in our sample presented as a red dot, suggesting a new upper limit that is slightly below the previously reported limit by \citet{2016ApJ...833L...8T}.} 
\label{fig:fig1}
\end{figure}

Figure \ref{fig:fig1} shows a scatter plot of the E$>$10 MeV peak proton flux versus  the X-ray flare intensity for the 65 SEP events in our sample (blue dots). The solid black line is the best-fit regression to the data in the log-log space. Similar to \citet{2012ApJ...756L..29C, 2013JSWSC...3A..31C}, and \citet{2019A&A...621A..67H} here we use the reduced major axis (RMA) method, while most commonly the ordinary least squares (OLS) regression is employed \citep[e.g.,][]{belov2005proton}. However, the RMA is specifically formulated to handle errors in both variables     \citep{harper2014reduced} and a non-causal relationship between the two variables is assumed \citep[see details in][]{till1973use}. As a result, $I_{P}$ $\propto$ $F_{SXR}^{\beta}$ with $\beta$=1.40$\pm$0.19 was retrieved. Additionally, the gray shaded envelope in Fig.~\ref{fig:fig1} provides the  error estimated while employing the fitting routine.\footnote{\url{https://docs.scipy.org/doc/scipy-0.19.0/reference/
generated/scipy.optimize.leastsq.html}} In detail, the Jacobian matrix is multiplied with the residual variances, estimated by the mean square errors. The resulting covariance matrix is then   used to derive the
standard error, and therefore the $\pm\sigma$ uncertainty obtained from the fit (correlation coefficient is $cc$=0.46). 
Based on \cite{2016ApJ...833L...8T}, the upper limit of the $I_{P}$ - $F_{SXR}$ relation is given by $I_{P}$ $\propto$ $F_{SXR}^{5/6}$ (dashed red line). This relationship is based on a chain of assumptions that brings together solar flares, CMEs, and SEPs. The starting point of this relation is the SXR flux  ($F_{SXR}$), which is the most commonly used index of flare magnitude. \cite{2016ApJ...833L...8T} assumed that F$_{SXR}$ is roughly proportional to the total energy released during flares (E$_{flare}$) (i.e., F$_{SXR} \propto$ E$_{flare}$). In addition, they argued that the kinetic energy of the CME ($E_{CME}$) is proportional to $E_{flare}$ and that the CME mass ($M_{CME}$) is the sum of mass within the gravitationally stratified AR. Finally, these authors further assumed that the total kinetic energy of solar energetic protons is proportional to $E_{flare}$ and that the duration of the proton flux enhancement is determined by the CME propagation timescale. As a result, the energetic proton flux $I_{P}$ in response to the SXR flare class ($F_{SXR}$) is scaled as

\begin{equation}\label{eq1}
 I_{P} \propto F_{SXR}^{5/6}   
.\end{equation}

\begin{figure*}[t!]
\centering
\includegraphics[width=\textwidth]{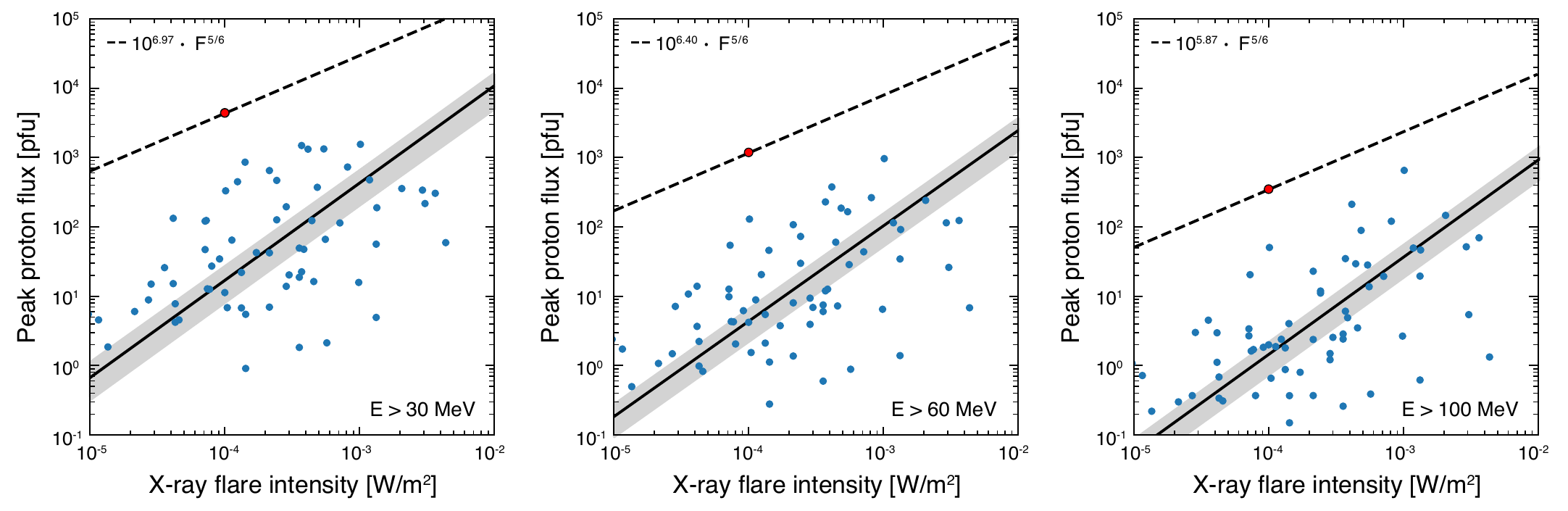}
\caption{Similar to Fig. \ref{fig:fig1}. From left to right, these panels present results for E$>$30-; E$>$60-; and E$>$100 MeV, respectively. The solid black line corresponds to the RMA regression in each case. The black dashed line gives the upper solar limit  based on the \cite{2016ApJ...833L...8T} scaling law of $I_{P}$ $\propto$ $F_{SXR}^{5/6}$. The line is forced to fit the uppermost point in each panel. In all panels, the red dot correspond to the 8 November 2000 outstanding large SEP event (see text for further details).} 
\label{fig:fig4}
\end{figure*}

As can be seen, there is only one SEP event in our sample that had an $I_{P}$ larger than 10$^{4}$ pfu (on 8 November 2000). This event was associated with an M7.0 SXR flare, and hence stands out in the plot as the central uppermost data value (red filled circle in Fig. \ref{fig:fig1}). This remarkable SEP event \citep[see][]{2019ApJ...877...11C} was associated with a well-connected source \citep[W77$^{o}$; e.g.,][]{lario2003solar}, a wide ($>$170$^{o}$) and fast CME \citep[$\sim$1700 km/s; see][]{thakur2016two}, a long-lasting type II radio burst \citep{agueda2012multi}, and a complex type III radio emission \citep{cane2002solar}. Although this SEP event had the potential to be registered by NMs and hence be listed as a GLE, there was no increase measured at ground-based detectors \citep{butikofer2021sep}. For this event, the height where the CME-driven shock formed, based on type II radio burst measurements, was estimated to be $\sim$ 3.5 $R_\odot$ \citep{thakur2016two}. This height is a factor of $\sim$2.3 above the median CME height for GLEs and is probably too high to accelerate GLE particles. We used the same theoretical arguments that were put forth by \citet{2016ApJ...833L...8T}, and thus the slope of the upper limit is kept identical. However, based on our sample, this upper limit had to be re-scaled, as indicated by the dashed black line. It should also be noted that the \citet{2016ApJ...833L...8T} sample is inclusive of the uppermost point in our sample (i.e., 8 November 2000). Their sample includes an even stronger event (on 4 November 2001) with a larger peak proton flux of 31700 pfu associated with an X1.0 SXR flare.\footnote{see \url{http://cdaw.gsfc.nasa.gov/CME_list/sepe/}} As we do in Figure \ref{fig:fig1}, \cite{2016ApJ...833L...8T} scale the theoretically derived $\propto F_{SXR}^{5/6}$ law in order to go through this extreme point. The 4 November 2001 event was not considered in our sample for two reasons:  because  the associated flare was located at W18$^\circ$, and is thus  outside our W20$^\circ$-W90$^\circ$ bin, and because  the obtained peak proton flux was related to the arrival of the CME-shock at Earth \citep{2008SoPh..252..409S}. Hence, the resulting equation of this line for our sample is a more realistic and yet conservative upper limit of $I_P = 10^{7.50}\cdot F_{SXR}^{5/6}$. 

\begin{table}[h!]
\centering
\caption[]{Slope of the $I_{P}$ -- $F_{SXR}$ relation and correlation coefficient (cc) for each integral energy derived in this work.}
\label{tab:table1}
\begin{tabular}{lcc}
\hline
\bf{Integral} & \bf{Slope} & \bf{Correlation}\\
 \bf{Energy}  & $I_{P}$ -- $F_{SXR}$ & \bf{Coefficient}\\
   \bf{[MeV]}          &   ($\beta$)         &     \bf{(cc)}      \\
\hline
E \textgreater{} 10       & 1.40$\pm$0.19 &0.39\\
E \textgreater{} 30       & 1.38$\pm$0.19 &0.44\\
E \textgreater{} 60       & 1.38$\pm$0.17 &0.49\\
E \textgreater{} 100      & 1.41$\pm$0.17 &0.52\\
\hline
\end{tabular}%
\end{table}

We then obtained similar solar scaling relations of the form $I_{P}\propto F_{SXR}^{\beta}$ for the integrated E$>$30 MeV, E$>$ 60 MeV, and E$>$ 100 MeV energy channels. Our results are presented in Fig. \ref{fig:fig4}, where the $I_P$--$F_{SXR}$ relations similar to Fig.~\ref{fig:fig1} for E$>$30 MeV (left panel), E$>$ 60 MeV (middle panel), and E$>$ 100 MeV (right panel) are displayed. Table~\ref{tab:table1} summarizes the slopes obtained by the RMA regression fits of each of the individual cases. 

The exponent (power-law index) $\beta$ seems to be relatively constant ($\sim$1.40) among the different energies, implying energy-independent slopes due, at least in part, to the interrelatedness of the four integral energies considered. The correlation coefficients of the $I_P$ -- $F_{SXR}$ relation seem to increase with energy. A somewhat similar trend has also been reported by \cite{2015SoPh..290..841D}. 

\subsection{From peak fluxes to fluences}
As described in Section \ref{sec:data}, the peak proton flux per integral energy and the fluence were calculated for each of the 65 SEP events under study (see Appendix \ref{appendix:C}). In the next step, scaling relations between fluences ($F_{P}$) and $I_{P}$ are derived. Although the scientific community routinely uses  $I_{P}$ values to associate SEP events with their parent solar events \citep{2010JGRA..115.8101C, 2016JSWSC...6A..42P, desai2016large}, the time--intensity profiles [$I_{(t)}$], resulting from the convolution of SEP acceleration and transport processes, are also needed. This is especially important when quantifying the radiation environment. Following the procedure discussed in \cite{kahler2018relating} we compared the measured fluences ($F_{P}$) and peak proton fluxes ($I_{P}$) for each integral energy investigated in this study. In agreement with \citet{kahler2018relating}, robust correlations ($cc$ $\approx$ 0.97) between $F_{P}$ and $I_{P}$ for each integral energy are found with slopes near unity. Thus, our results indicate energy independence in the relationship between $I_{P}$ and $F_{P}$. The details of these relations and the corresponding validation and verification comparisons are discussed in Appendix \ref{appendix:A}.

Once the fluences ($F_{P}$) were derived from the data, we obtained the corresponding relations between $F_{P}$ $\propto$ $F_{SXR}$ by employing the RMA regression method (similar to the relations discussed in Section \ref{sec:1}). The derived slopes $F_{P}$ -- $F_{SXR}$ and correlation coefficients are presented in Table \ref{tab:flu}. The correlation coefficients show a relative increase with respect to the increasing integral energy, starting at 0.43 for E$>$10 MeV and reaching 0.54 for E$>$100 MeV.

\begin{table}[h!]
\centering
\caption[]{Slope of $F_{P}$-$F_{SXR}$ and correlation coefficient ($cc$) for each integral energy derived in this work.}
\label{tab:flu}
\begin{tabular}{lcc}
\hline
\bf{Integral} & \bf{Slope} & \bf{Correlation}\\
 \bf{Energy}  & $F_{P}$ -- $F_{SXR}$ & \bf{Coefficient}\\
   \bf{[MeV]}          &   ($\delta$)         &     \bf{(cc)}      \\
\hline
E \textgreater{} 10       & 1.59$\pm$0.21 & 0.43\\
E \textgreater{} 30       & 1.56$\pm$0.20 & 0.48 \\
E \textgreater{} 60       & 1.48$\pm$0.18 & 0.52\\
E \textgreater{} 100      & 1.44$\pm$0.17 & 0.54\\
\hline
\end{tabular}%
\end{table}


\subsection{Integral energy spectra for the sample events}
For each of the 65 SEP events in our sample,  we fit the derived peak fluxes and fluences at the four integral energies under consideration with an inverse power law (A$^{-\epsilon}$). We then compute a mean and median spectrum (i.e., $I_{P}$ vs. $>E$ and $F_{P}$ vs. $>E$) as follows: (a)  determining the sample mean and median values of $I_{P}$ and $F_{P}$ at each of the four energies based on the fitted values of these parameters for the 65 events; (b)  fitting these values under the same assumption of inverse power-law dependence.
 
\begin{figure}[t!]
\centering
\includegraphics[width=\columnwidth]{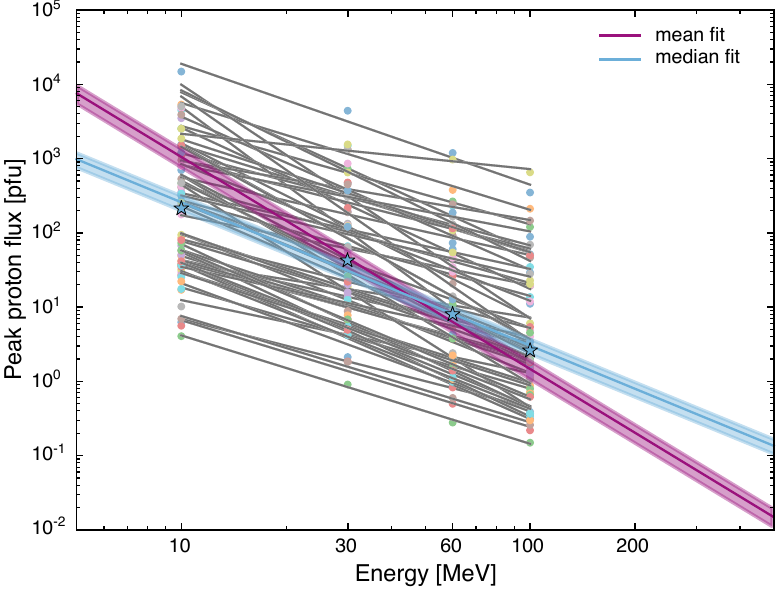}
\caption{Derived spectrum for the 65 SEP events, assumed to follow an inverse power law (gray lines). The mean spectrum is shown as a continuous magenta line ($\epsilon_{mean}$ = 2.87), while the blue line provides the median spectrum ($\epsilon_{median}$= 1.96). The blue stars represent the median peak proton flux values per energy.} 
\label{fig:spectrum}
\end{figure}

The results for the integral spectra of peak proton fluxes are presented in Fig.~\ref{fig:spectrum}. Here the colored dots represent the measured $I_{P}$ values of each SEP event, while the solid gray lines show the event-dependent inverse power-law fits. In addition, the solid purple and blue lines represent the mean ($\epsilon_{mean}$=2.87) and median ($\epsilon_{median}$=1.96) spectra derived from our SEP sample, respectively.     

\section{Estimating peak proton flux and fluence for solar superflares}
\label{sec4}

\subsection{Estimates and arguments}
In this section we attempt to estimate the uppermost peak proton fluxes ($I_{P}$) and fluences ($F_{P}$) for two notable SEP parent flares. Specifically, we focus on the event of 23 February 1956 (GLE05) \citep{2005AdSpR..35..697B}, the most intense high-energy SEP event of the modern era,  and the AD774/775 SEP event \citep{2020ApJ...903...41C}.
The radionuclide records show peak-like increases  on the order of 12\textperthousand\,\, around AD774/775. Since no information on the corresponding SEP spectrum is known for such an event, a scaling of GLE05 is usually assumed, with a multiplicative value of 70$\pm$30 applied to the 1956 SEP spectrum to obtain that of AD774/775 \citep[see Table 1 of][]{usoskin2021strongest}.

The SEP-to-SXR flare scaling law of \cite{2016ApJ...833L...8T} in Eq. (\ref{eq1}) is shown as dashed black lines in Figs. \ref{fig:fig1} and \ref{fig:fig4}. In each plot the relations are positioned to run through the most intense SEP events of our sample so that we can discuss the upper limits of the peak proton flux ($I_{P, upper}$) in response to $F_{SXR}$. Based on our sample %
\begin{equation} \label{eq3}
 I_{P, upper} = I_{P, energy} \cdot F_{SXR}^{5/6},
\end{equation}
where $I_{P, E10} = 10^{7.50}$ pfu, $I_{P, E30} = 10^{6.97}$ pfu, $I_{P, E60} = 10^{6.40}$ pfu, and $I_{P, E100} = 10^{5.87}$ pfu, and where $F_{SXR}$ is normalized in units of 1 $W/m^{2}$. \\ 
\begin{figure*}[!t]
\centering
\includegraphics[width=\textwidth]{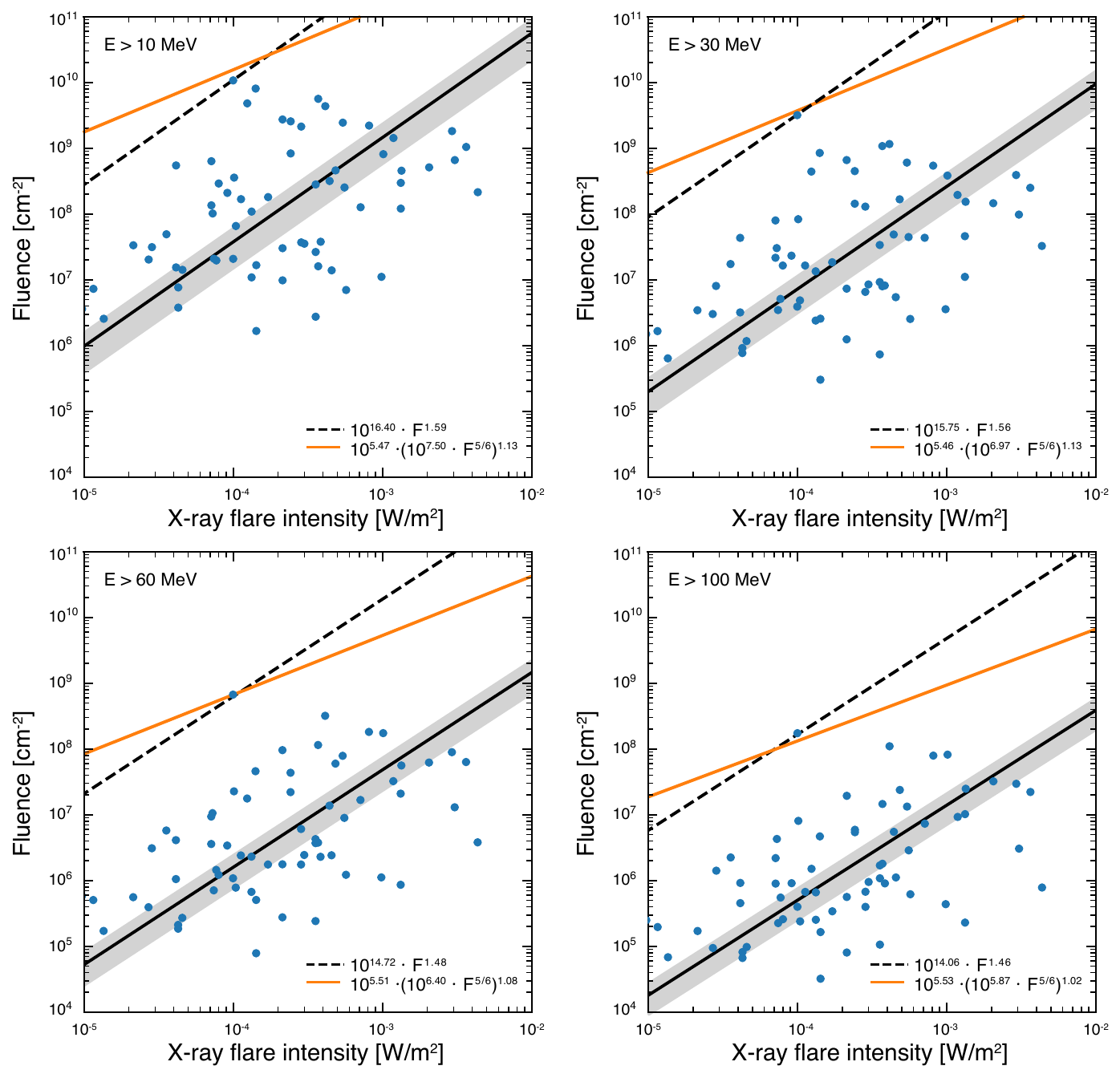}
\caption{$F_{P} \propto F_{SXR}$ relations for the four integral energy bands of the SEP events:  E$>$10 (top left); E$>$30 (top right); E$>$60 (bottom left); and E$>$100 MeV (bottom right). The log-log relations are obtained with the RMA regression fitting (solid black line). The estimated upper limits of $F_{P}$ in terms of $F_{SXR}$ based on \cite{2016ApJ...833L...8T} are depicted as solid orange lines in each panel. The dashed black lines are   similar to the RMA line scaled to the uppermost point of the sample.} 
\label{fig:fig7}
\end{figure*}
%
\begin{figure*}[h!]
\centering
\includegraphics[width=\textwidth]{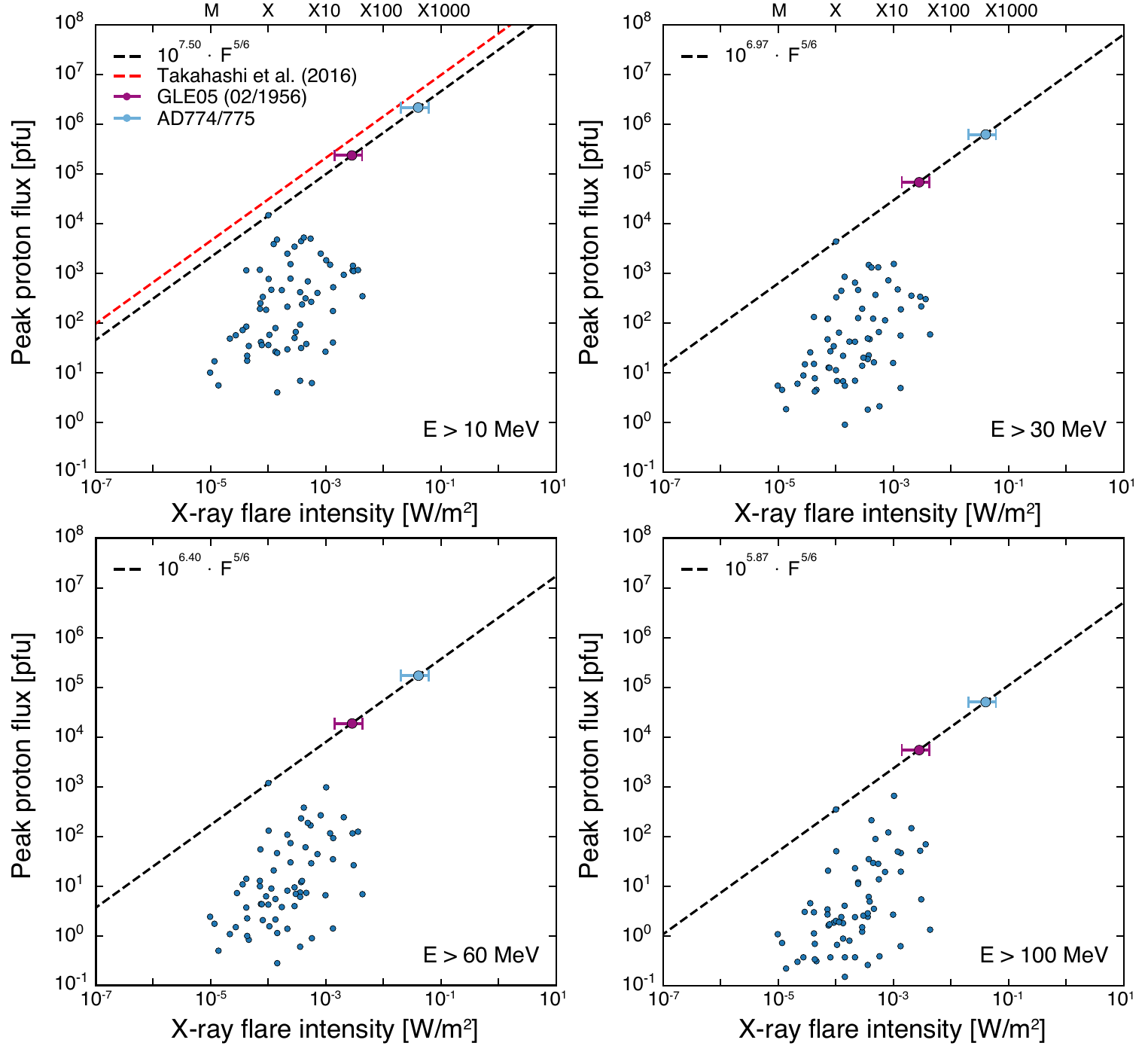}
\caption{$I_{P} \propto F_{SXR}$ relations for the four integral energy bands of the SEP events:  E$>$10 (top left); E$>$30 (top right); E$>$60 (bottom left); and E$>$100 MeV (bottom right). The estimated SXR flare mean values for the events of 23 February 1956 and AD774/775 are represented as purple and cyan filled circles, respectively. The plotted SXR intensities were obtained by dividing the values in Appendix \ref{appendix:C} by 0.7 to reflect the recent NOAA recalibration \citep[Hudson et al. in preparation;][]{cliver2022}.} 
\label{fig:fig6}
\end{figure*}
%
Combining the peak proton flux and fluence relations from Appendix \ref{appendix:A} (see Fig.~ \ref{fig:fig5}) with the fits from the dashed lines in Figs.~\ref{fig:fig1} and \ref{fig:fig4}, the upper-limit fluences $F_{P, upper}$ are found to be best described by

\begin{equation} \label{eq5}
 F_{P, upper} = F_{P, energy} \cdot (F_{SXR}^{5/6})^{\delta},   
\end{equation}

\noindent where $F_{P, E10} = 10^{5.48}$ $\cdot$ $(10^{7.50})^{\delta}$ cm$^{-2}$, $F_{P, E30} = 10^{5.46}$ $\cdot$ $(10^{6.97})^{\delta}$cm$^{-2}$, $F_{P, E60} = 10^{5.51}$ $\cdot$ $(10^{6.40})^{\delta}$ cm$^{-2}$, and $F_{P, E100} = 10^{5.53}$ $\cdot$ $(10^{5.87})^{\delta}$ cm$^{-2}$, and where  $F_{SXR}$ is normalized in units of 1 $W/m^{2}$. The values of $\delta$ can be found in Appendix \ref{appendix:A} (see Table \ref{tab:fluence}). These results are used in Fig.~\ref{fig:fig7} to show the $F_{P}$ -- $F_{SXR}$ upper limit relations for each integral energy   (orange line).\\

The solid black lines shown in Fig.~\ref{fig:fig7} correspond to the RMA regression applied to the 65 SEP events (see Table \ref{tab:flu}) embedded in the error band (gray shaded area), while the relationship in Eq.~(\ref{eq5}) is represented by orange lines, and the dashed black lines present an alternative upper limit based on the RMA fit shifted to fit the uppermost $F_{P}$ values of the SEP sample.

\subsection{Solar flares in February 1956 and AD774/775}
The 23 February 1956 GLE event (GLE05) is the most extreme GLE event yet recorded, with a similar structure to the 20 January 2005 event (GLE69). The strongest solar flare in February 1956 occurred when the AR group 17351 was near the west limb as seen from Earth. At this time a solar flare of H$\alpha$ importance class 3B, located at N25 W85, took place at 03:34 UT and produced this notable GLE \citep{belov2005solar}. The fluence of GLE05 at an integral energy of E $>$ 430 MeV ($>$ 1 GV in rigidity) was calculated to be 4.21$\times$ 10$^7$ cm$^{-2}$ \citep{usoskin2020revised} being almost one order of magnitude higher than the fluence of any other known GLE \citep[see Table 1 of][]{2020ApJ...903...41C}, making it the largest SEP event ever recorded by modern instrumentation. 

By examination of concentrations of cosmogenic nuclides sequestered in tree rings and ice cores, intense SEP events have been identified in the distant past. The AD774/775 event was the first such event discovered as an exceptional increase in $^{14}$C concentration (12\textperthousand)  in tree rings \citep{miyake2012signature}. This increase was investigated independently in detail by many different groups \citep[e.g.,][]{usoskin2013ad775,mekhaldi2015multiradionuclide,2018NatCo...9.3605B} to  verify its solar origin \citep{mekhaldi2015multiradionuclide}. As a result, this exceptional SEP event was recently modeled by a SEP fluence spectrum $\sim$ 62 times that of GLE05 \citep[][see their Table 1]{2020ApJ...903...41C}, with a a E$>$430 MeV fluence of 1.6 $\times 10^9$ cm$^{-2}$, making it one of the most powerful inferred SEP events to date. This corresponds to the consensus value of several of these independent studies of $^{14}$C and $^{10}$Be concentrations in tree rings and ice cores. More such SEP superevents have been found in the cosmogenic radionuclide records: AD993/994 \citep{Miyake-etal-2019}, 660 BC \citep{O'hare-etal-2019}, and most recently 5410 BC \citep[][]{Miyake-etal-2021}, 7176 BC and 5259 BC \citep[both discussed in][]{Brehm-etal-2021}, and 9125 BP \citep[][]{Paleari-etal-2022}, but thus far no estimates of the corresponding SXR flare class have been made for these events.

\cite{2020ApJ...903...41C} estimated that $F_{SXR}$ associated with the 23 February 1956 GLE event ranged from X10 to X30, and from this estimate obtained a SXR class of X145 to X425 for the AD774/775 flare. In particular, \citet{2020ApJ...903...41C}; their Figure 7)  applied an RMA fit to a scatter plot of modeled E$>$200 MeV fluences from \cite{2018JSWSC...8A...4R} for hard-spectrum GLEs versus the peak intensities of their associated SXR flares.  They then add two points for the 1956 GLE (i.e., GLE05) based on the estimated range of the peak SXR intensity (X10-X30) of its associated flare (as inferred from white-light, radio, sudden ionospheric disturbances, comprehensive flare index, inferred CME transit time, and geomagnetic storm observations) and its E$>$200 MeV fluence \citep{Usoskin_etal_2020}. Through these points they extrapolated lines parallel to the RMA fit to the modeled E$>$200 MeV fluence for the AD774 SEP event to obtain an estimate of X285$\pm$140 for the AD774 flare. Based on the NOAA reassessment of the recent SXR calibration, these values were corrected to X14 to X42 (GLE05) and X200 to X600 (AD774/775), respectively \citep[see][]{cliver2022}. Following Eq.~(1) of \cite{2020ApJ...903...41C}, corrected for the SXR flux shift\footnote{$F_{TSI}=0.33 \rm x 10^{32}(C_{GOES}/C_{GOES, X1.4})^{0.72}$, where   $F_{TSI}$ is the flare total solar irradiance of bolometric energy and $C_{GOES}/C_{GOES, X1.4}$ equals the flare SXR class scaled to X1.4.}, the flare bolometric energy for the upper limits of the SXR associated flares for the nominal values of each of these SEP events is currently  $\sim$3 $\times$ 10$^{32}$ erg (1956) and $\sim$2 $\times$ 10$^{33}$ erg (AD774/775). For comparison, radiative energies of 3.6 $\times$ 10$^{32}$ erg and  4.3 $\times$ 10$^{32}$ erg were recorded for $>$X10 flares on  28 October and  4 November 2003, respectively \citep[from TIM measurements;][]{2012ApJ...759...71E}.

\begin{figure*}[t!]
\centering
\includegraphics[width=\textwidth]{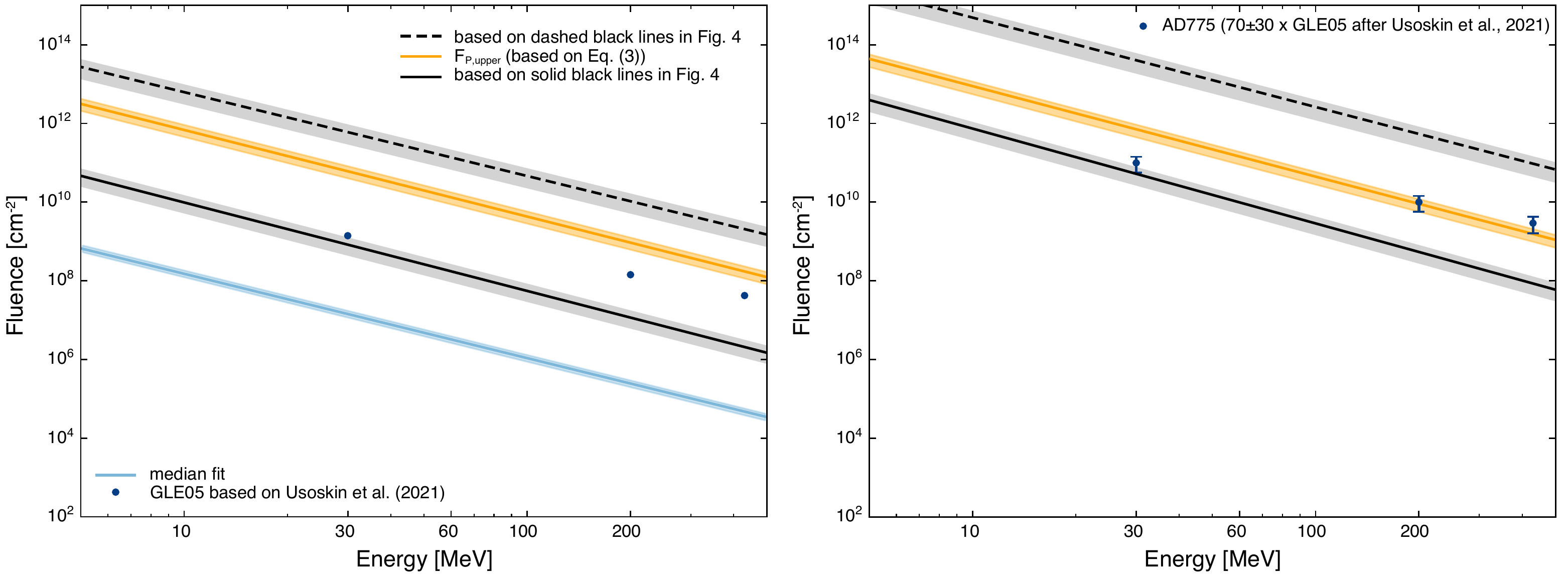}
\caption{Derived fluence spectra for the two solar cases under investigation. The left panel corresponds to GLE05;  the right panel corresponds to the AD774/775 SEP event. These plots are similar to Fig.~\ref{fig:spectrum}, but for a fluence with the addition of the relevant data points for each event (based on \citealt{usoskin2021strongest}), depicted as filled blue circles, and the corresponding derived spectra calculated for each case from the relations presented in Sect.~\ref{sec:scaling}. The median spectra is shown (light blue line) with the corresponding error (gray shade).}
\label{fig:fig8}
\end{figure*}
%
\subsection{SEP peak fluxes ($I_{P}$) and fluences ($F_{P}$) driven by $F_{SXR}$}

Figure \ref{fig:fig6} shows the $I_{P}$ -- $F_{SXR}$ relations for the four respective integral energy bands. The estimated peak proton fluxes of GLE05 and AD774/775 are indicated by the purple and blue filled circles, respectively. They are based on the mean SXR fluxes provided by \citet{2020ApJ...903...41C}, adjusted upward by  a factor of 1/0.7; $\sim$40\% for the NOAA recalibration (Hudson et al. 2022; in preparation), and range from X28$\pm$14 (GLE05) and $\sim$X400$\pm$200 (AD774/775). Substituting the corresponding $F_{SXR}$ in Eq.~(\ref{eq3}) then leads to the expected upper peak proton flux limits. The dashed black lines further present the upper limits given by Eq.~(\ref{eq3}), whose upper limit $I_{P}$ for GLE05 (based on $F_{SXR}$ = X42) is $I_{P, E>10}$ = 3.31 $\times$ 10$^{5}$ pfu, $I_{P, E>30}$ = 9.76 $\times$ 10$^{4}$ pfu, $I_{P, E>60}$ = 2.63 $\times$ 10$^{4}$ pfu, and $I_{P, E>100}$ = 7.75 $\times$ 10$^{3}$ pfu. For the AD 774/775 event (based on $F_{SXR}$=X600) this would be $I_{P, E>10}$ = 3.03 $\times$ 10$^{6}$ pfu, $I_{P, E>30}$ = 8.95 $\times$ 10$^{5}$ pfu, $I_{P, E>60}$ = 2.41 $\times$ 10$^{5}$ pfu, and $I_{P, E>100}$ = 7.11 $\times$ 10$^{4}$ pfu (see also Table \ref{tab:res1}).

\begin{figure}[t!]
\centering
\includegraphics[width=\columnwidth]{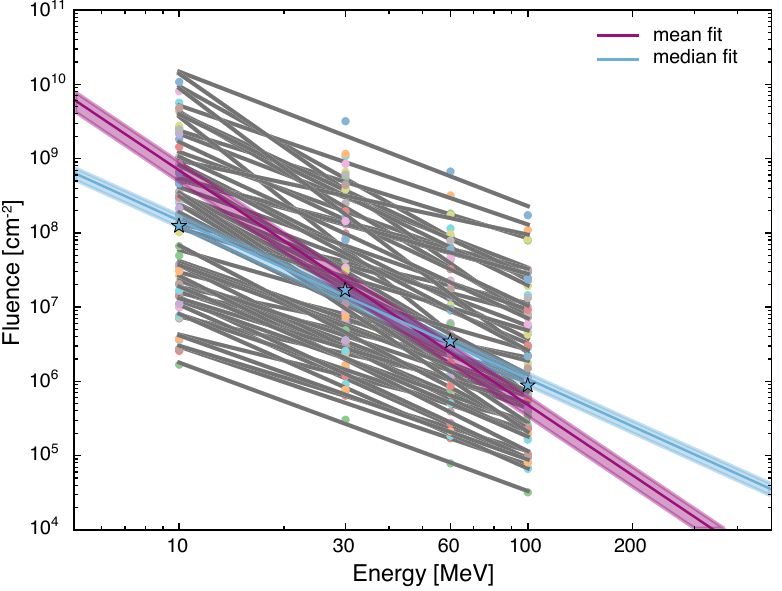}
\caption{Similar to Fig.~\ref{fig:spectrum}, but for fluence. The derived fluence spectrum for each of the 65 SEP events is presented as a solid black line. The mean spectrum is shown as a continuous magenta line ($\epsilon_{mean}$=3.16), while the blue line provides the median spectrum ($\epsilon_{median}$=2.13). The blue stars represent the median fluence values per energy.}
\label{fig:fig12a}
\end{figure}

\begin{table}[h!]
\caption[]{Upper limit peak proton fluxes ($I_{P}$, [pfu]) and fluences ($F_{P}$, [cm$^{-2}$]) for the SEP event on AD774/775 and GLE05 derived in this work, for each integral proton energy. Peak proton fluxes were calculated via Eq.~\ref{eq3} and fluence via Eq.~\ref{eq5} for a given SXR flux. The upper and lower limits listed here are driven by the $F_{SXR}$ range of the associated solar flare per event, and are from \cite{cliver2022}.}
\label{tab:res1}
\begin{tabular}{lcc}
\hline
             & \bf{AD774/775} & \bf{GLE05}\\
             \hline
\bf{Integral Energy} & \multicolumn{2}{c}{\bf{Peak Proton Flux -- $I_{P}$}} \\
  \bf{(MeV)} & \bf{(pfu)} & \bf{(pfu)}\\
\hline
\\[-0.5em]
E \textgreater{} 10       & 2.16E+06$_{1.21 \rm E+06}^{3.03\rm E+06}$ & 2.36E+05$_{1.32\rm E+05}^{3.31 \rm E+05}$\\
\\[-0.5em]
E \textgreater{} 30       & 6.38E+05$_{3.58\rm E+05}^{8.95\rm E+05}$ & 6.96E+04$_{3.91\rm E+04}^{9.76\rm E+04}$ \\
\\[-0.5em]
E \textgreater{} 60       & 1.72E+05$_{9.64\rm E+04}^{2.41\rm E+05}$ & 1.87E+04$_{1.05\rm E+04}^{2.63\rm E+04}$\\
\\[-0.5em]
E \textgreater{} 100      & 5.07E+04$_{2.85\rm E+04}^{7.11\rm E+04}$ & 5.53E+03$_{3.10\rm E+03}^{7.75\rm E+03}$\\
\\[-0.5em]
\hline
\bf{Integral Energy} & \multicolumn{2}{c}{\bf{Fluence -- $F_{P}$}} \\
  \bf{(MeV)} & \bf{(cm$^{-2}$)} & \bf{(cm$^{-2}$)}\\
\hline
\\[-0.5em]
E \textgreater{} 10       & 4.25E+12$_{2.21\rm E+12}^{6.23\rm E+12}$ & 3.48E+11$_{1.81\rm E+11}^{5.09\rm E+11}$\\
\\[-0.5em]
E \textgreater{} 30       & 1.05E+12$_{5.45\rm E+11}^{1.53\rm E+12}$ & 8.56E+10$_{4.45\rm E+10}^{1.25\rm E+11}$ \\
\\[-0.5em]
E \textgreater{} 60       & 1.46E+11$_{7.82\rm E+10}^{2.10\rm E+11}$ & 1.33E+10$_{7.14\rm E+09}^{1.92\rm E+10}$\\
\\[-0.5em]
E \textgreater{} 100      & 2.13E+10$_{1.18\rm E+10}^{3.01\rm E+10}$ & 2.23E+09$_{1.24\rm E+09}^{3.14\rm E+09}$\\
\\[-0.5em]
\hline
\end{tabular}%
\end{table}


Following the same reasoning, the upper limit fluences ($F_{P}$) in terms of $F_{SXR}$ for each event were then calculated utilizing Eq.~(\ref{eq5}), with the relevant $F_{P, energy}$ per case. These upper limits are given as orange lines in each of the panels of Fig.~\ref{fig:fig7}. A summary of the results for both the upper limit peak proton flux and fluence at each integral energy of interest for GLE05 and the AD774/775 event are presented in Table \ref{tab:res1}. 

\subsection{Spectrum based on $F_{SXR}$}
The dark blue circles in Figure \ref{fig:fig8} give the fluence spectra for the February 1956 (left panel) and AD774/775 (right panel) SEP events. The blue circles in the left panel are estimates of the fluence of GLE05 given by \citet{usoskin2020revised}, while those for the AD774/775 event shown in the right panel are scaled by a factor of 70$\pm$30 \citep[][indicated by the error bars]{usoskin2020revised}. Both panels in Figure \ref{fig:fig8} contain four lines (three of which are derived from similarly formatted lines, black dashed, orange, black solid) in Figure \ref{fig:fig7}, and a fourth line (light blue)  taken directly  from Figure \ref{fig:fig12a}. The three lines in Figure \ref{fig:fig8} that stem from Figure \ref{fig:fig7} are all constructed in the same way; the fluences are obtained from the corresponding lines in each of the  four energy panels in Figure \ref{fig:fig7} for a SXR class of X42 (1956) and $\sim$X600 (AD774/775) for the left and right panels, respectively, in Figure \ref{fig:fig8}. For each line the shaded areas provide the error.

The obtained integral fluence spectra depicted by the dashed black, solid black, and solid orange lines in Fig.~\ref{fig:fig8} are driven by the associated $F_{SXR}$, in this case an $\sim$X600 class flare, corresponding to the upper limit range of the AD774/775 superflare, based on a 62 times multiple of the 1956 SEP spectrum \citep{2020ApJ...903...41C}. The plotted dark blue points for the slightly higher multiple of 70 from \citet{usoskin2020revised} used to scale the AD774/775 SEP event to the 1956 GLE fall below  the orange upper limit line (or within
the uncertainty) at E$>$430 MeV) based on the \cite{2016ApJ...833L...8T} scaling relationship ($F_{P} - F_{SXR}^{5/6}$) in Figure \ref{fig:fig8}. The proximity of the E$>$200 MeV and E$>$430 MeV points to the orange constraint line suggests that the AD774/775 event is close to the limit of what the Sun is capable of producing for such an $F_{SXR}$, as has been surmised by others \citep[see Figure 7.3 in][]{Miyake-etal-2019}.
 

\subsection{Consideration of additional scaling relationships}
In this section we consider two additional scaling laws. In the first  we take into account a longer acceleration process within the inner heliosphere and assume that the peak proton flux ($I_P$) is proportional to $E_{flare} \cdot V_{CME}$. Combining $E_{flare} \propto F_{SXR}$ with $V_{CME} \propto F_{SXR}^{1/6}$ \citep[see][]{2016ApJ...833L...8T} a scaling law of the form $F_{SXR}^{7/6}$ is derived (versus the $F_{SXR}^{5/6}$ in Eq.~(\ref{eq1})). In the second we consider an intermediate scaling law in which direct proportionality of the peak proton flux ($I_P$) to the total number of accelerated particles at the energy under consideration is assumed, and thus $I_P \propto E_{flare} \propto F_{SXR}$. A scaling law of $F_{SXR}$ practically leads to no difference at all for each of the integral energies employed in this work. On the other hand a scaling of $F_{SXR}^{7/6}$ leads to larger peak proton fluxes for stronger solar flares (i.e., larger $F_{SXR}$).

Figure \ref{fig:DISCUSSION} shows the evaluation of the scaling relation with $F_{SXR}$ (on the left) and $F_{SXR}^{7/6}$ (on the right) against the scaling with $F_{SXR}^{5/6}$ in the case of the obtained fluences for the two events under consideration, namely GLE05 (top panels) and AD774/775 (bottom panels). As a result, for the first comparison  (i.e., $F_{SXR}^{5/6}$~vs.~$F_{SXR}$) it seems that there is a marginal difference and in particular, we calculated that there is a mean relative factor, covering a wide energy range from E$>$10 to E$>$500 MeV, of 1.35 (2.13) for the case of GLE05 (AD774/775). For the second comparison (i.e., $F_{SXR}^{5/6}$~vs.~$F_{SXR}^{7/6}$) it seems that there is a difference of a factor of $\sim$12 for lower energies (i.e., E$>$10 MeV), but the higher the energy, the better the agreement between the scalings. Following the same procedure as before, the calculated mean relative factor for the same wide energy is 10.5 (6.64), for GLE05(AD774/775), respectively. More importantly,   all three possible scaling laws that are based on theoretical arguments, downstream of the \cite{2016ApJ...833L...8T}, offer a range of the obtained fluence for high-energy particles (i.e., E$>$200 MeV) that are in agreement with each other within a factor of $\sim$3.

\section{Discussion and conclusions}
\label{sec:conclusion}
In this work, scaling relations of the peak proton flux ($I_{P}$) and the fluence ($F_{P}$) of large well-connected SEP events recorded at Earth between 1984 and 2017, to the SXR peak fluxes ($F_{SXR}$) of their associated flares were investigated. In contrast to previous studies \citep[e.g.,][]{2016ApJ...833L...8T}, the scaling relations discussed in this work (i.e., $I_{P}$ or $F_{P}$ vs. $F_{SXR}$) are not limited to the integral proton energy of E$>$10 MeV, but have also been calculated for a broader set of integral energies: E$>$30; E$>$60; and E$>$100 MeV.

In the case of the $I_{P} \propto F_{SXR}^{\beta}$ relations, we found that the power-law index $\beta$ seems to be almost constant ($\sim$1.40) with increasing energy. This is consistent with the result of \cite{2007SoPh..246..457B}, who used a different sample and employed the OLS method (instead of the RMA method used in this work) and found that the slope for the $I_{P} \propto F_{SXR}$ relation is practically constant for E$>$10 MeV and E$>$100 MeV ($\sim$ 0.95). The results argue in favor of a relation between X-ray and proton emissions within uncertainties, without excluding other eruptive manifestations, for example  coronal mass ejections (CMEs). In particular, \cite{2016ApJ...833L...8T} invoked proportional scaling laws among total flare energy ($E_{flare}$), CME kinetic energy, and the total SEP energy ($E_{SEP}$) to further derive a scaling of $I_P \propto V_{CME}^{5}$, assuming that the SEP event duration is inversely proportional to $V_{CME}$. Finally, calculations of the omnidirectional $F_{P}$ values allowed for an investigation of the relations between $I_{P}$ and $F_{P}$ (see Appendix \ref{appendix:A} for details). 

\begin{figure*}[t!]
\centering
\includegraphics[width=0.49\textwidth]{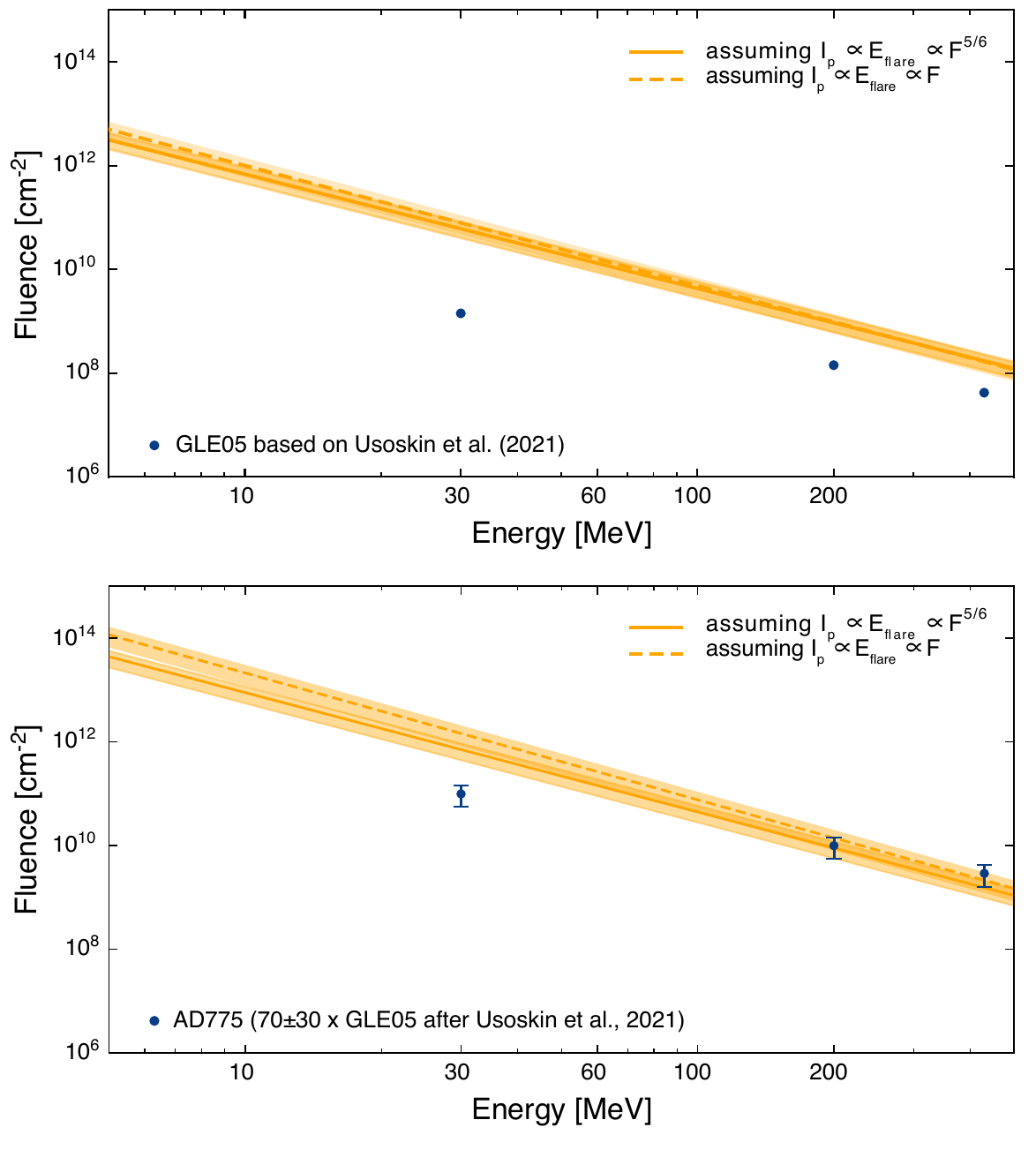}
\includegraphics[width=0.49\textwidth]{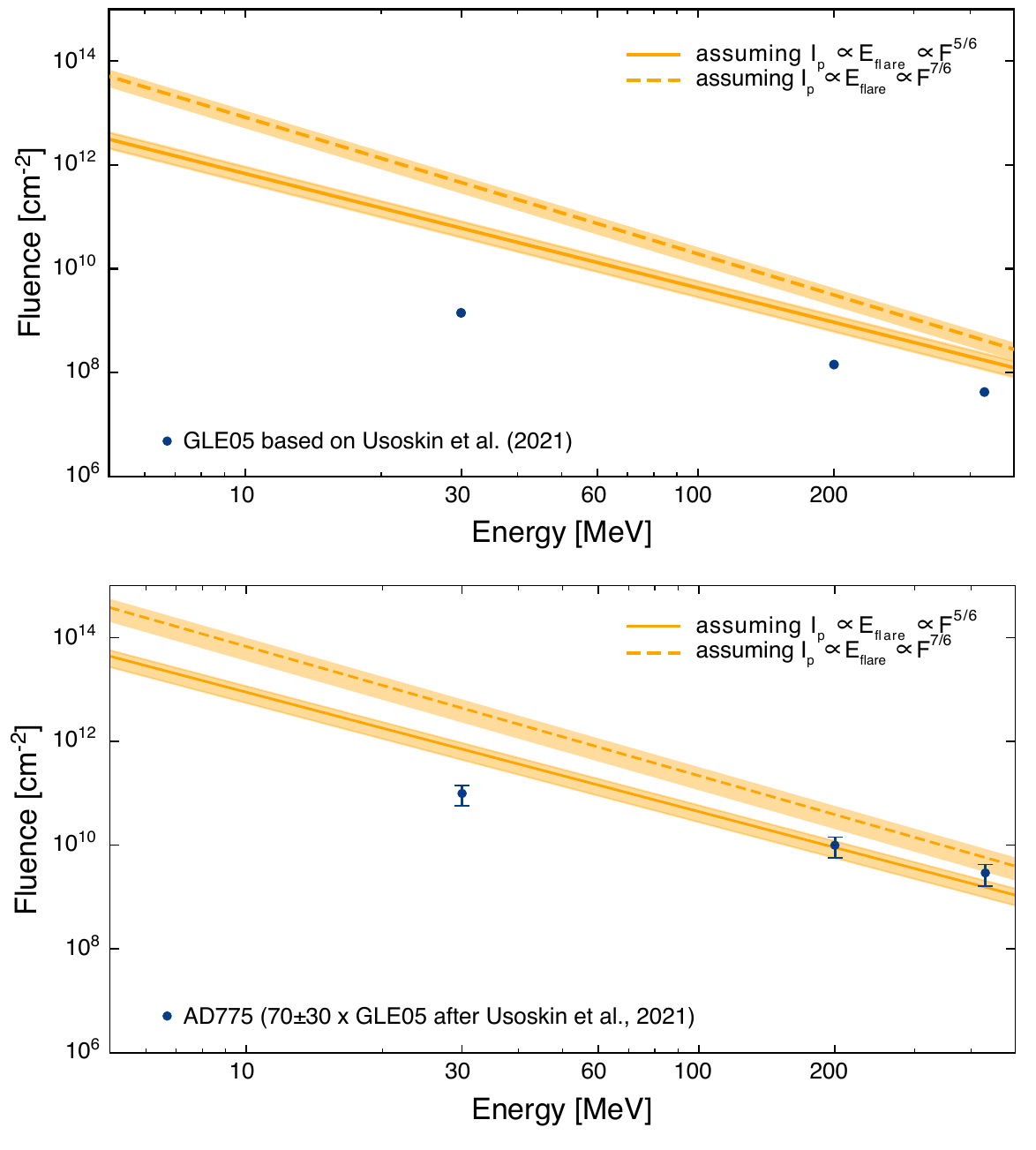}
\caption{Same as  Fig. \ref{fig:fig8}, but  investigating   different scaling applications. Left panels: $I_P \propto E_{flare} \propto F_{SXR}^{5/6}$ (solid lines) vs. $I_P \propto E_{flare} \propto F_{SXR}$ (dashed lines). Right panels: $I_P \propto E_{flare} \propto F_{SXR}^{5/6}$ (solid lines) vs. $I_P \propto E_{flare} \propto F_{SXR}^{7/6}$ (dashed lines). The results for GLE05 are shown in the upper panels, those for AD774/775 in the lower panels.}
\label{fig:DISCUSSION}
\end{figure*}

The upper limits of $I_{P}$ and $F_{P}$ were explored with the application of the $F_{SXR}^{5/6}$ scaling relations (Eq.~(\ref{eq3})) for $I_{P}$, as proposed by \cite{2016ApJ...833L...8T} for the case of E$>$10 MeV alone. In this work we showed that the $I_P \propto F_{SXR}^{5/6}$ scaling relations could also be successfully translated to fluence ($F_{P}$) scaling relations (see Eq.~(\ref{eq5})). Furthermore, the obtained fluences and integral spectra seem to represent quite reasonably the fluences derived independently by other studies (e.g.,  \citealt{usoskin2020revised}) of the AD774/775 event within the uncertainty limits (orange line and green filled circles in Fig.~\ref{fig:fig8}). Moreover, we obtained similar results for the two alternative scaling factors   shown in Figure \ref{fig:DISCUSSION}.


The scaling relations presented in this study provide a direct estimation of the upper limit peak proton flux ($I_{P}$), fluence ($F_{P}$), and spectrum ($\propto F_{SXR}^{5/6}$) based on the $F_{SXR}$ flux of the driving solar flare. Therefore, the relations allow us to  quantify the effect of such a flare on the radiation environment, within the uncertainties, caveats, and assumptions. For instance, such scaling laws are inherently based on the assumption that the SEP events result from a scaling of the flare energy producing the SXR flares \citep[e.g.,][]{2012ApJ...759...71E}. This is highlighted by several studies across decades of research \citep[e.g.,][]{1978SoPh...57..237H, belov2005solar, 2007SoPh..246..457B, 2012ApJ...756L..29C, 2019A&A...621A..67H} and suggests a causal relation between the solar eruption and the SEP production. Nonetheless, a recent   work argues against a close physical relation of solar flares and SEPs \citep{2013ApJ...769...35K} ignoring the evidence that flares associated with SEPs fundamentally differ from ordinary flares \citep{2007SoPh..246..457B}. Thus, our work begins with a sample of SEP events addressing the valid concern risen by \citet{2013ApJ...769...35K} that large SEP events do not arise in confined flares, making any general correlation between all flares and such SEP events problematic. CMEs cannot be excluded from such an approach of scaling relations. The work of \cite{2012ApJ...756L..29C} showed that scaling relations of SEPs do take into account CMEs. This is because flares associated with fast CMEs ($V_{CME}> 1000 km/s$) lead to  scaling laws that are similar to  those obtained for flares associated with SEPs. In addition, the correlation of fast CMEs to gradual SEPs have long been known \citep[e.g.,][and references therein]{2001JGR...10620947K}, while the works of \cite{2017Ge&Ae..57..727B} and \cite{2016ApJ...833L...8T} further quantifies this. Recently \cite{2020ApJ...901...63K} showed that flares with no CMEs are quite similar to flares associated with slow CMEs, in contrast to those flares associated with fast CMEs; suggesting the possibility of two classes of flares. These issues are noted here, and we should note that the scaling behavior of $I_P$ to $V_{CME}$ will be addressed in the second part of our study (Part II). 

Another issue to take into account is the dependence of the obtained upper limit presented in this work on the 8 November 2000 SEP event. Thus, we further assessed the possibility to exclude this event and re-applied our proposed methodology. We found a difference of one order of magnitude between the upper-limit spectra (including the 8 November 2000 SEP event) and the spectra obtained excluding this event. In turn, this means that the fluence ($F_P$) is about one order of magnitude lower than that obtained from the upper-limit bound, for the corresponding $F_{SXR}$, when taking into account the 8 November 2000 event. However, when excluding this SEP event the obtained spectra for GLE05 underestimates the actual measured fluence at E$>$200 MeV and primarily at E$>$430 MeV, while for the AD774 SEP event the underestimation  already starts at E$>$30 MeV (not shown), implying that the exclusion of this event underestimates the obtained fluence spectra at both cases. Scalings with $F_{SXR}$ and $F_{SXR}^{7/6}$ were also applied and showed similar underestimates of the derived fluence in both cases. In addition to these findings, to address the possible influence that the inclusion of the event on 8 November 2000 has on our calculations, we calculated the residuals of the relations $I_P$ to $F_{SXR}$ for each integral energy (see Figures~\ref{fig:fig1} and \ref{fig:fig4}) and showed that although there is a considerable spread for all 65 points, the residual of 8 November 2000 falls well within 3 standard deviations of the residuals ($\rm  S.D.$) (see details in Appendix \ref{appendix:B}). Therefore, this event was not excluded from the analysis.

Differences in the obtained scaling laws of $I_P$ distributions for flares alone and SEP associated flares have been a general point of intensive research since \cite{1978SoPh...57..237H}. According to \cite{1978SoPh...57..237H} as also recently summarized by \cite{2020ApJ...901...63K}, there are three possibilities to account for such a discrepancy: (a) flares associated with SEPs are fundamentally different from ordinary flares; (b) flares associated with SEPs represent the high end of the energy distribution of ordinary flares;  (c) flares associated with SEPs exceed a threshold barrier. \cite{2018ApJ...864...48C} argue in favor of (a) and (c), indicating that this threshold is at a CME speed of $\sim$400 km/s, above which a shock can be formed \citep[see Figure 2 of][]{2018ApJ...864...48C}. This is further supported by the fact that fast CMEs are needed in order to drive shocks capable of accelerating SEPs. Moreover, this is also corroborated by \cite{2020ApJ...901...63K} who showed that flares associated with SEP events and/or fast CMEs are characterized by lower flare temperatures than those without. On the other hand, (b) is favored by the work of \cite{2016ApJ...833L...8T}, although \cite{2013ApJ...769...35K} argued against the validity of scaling laws in general. Nonetheless, scaling relations hold important information providing a representation of the expected conditions of the radiation environment. 

An important point in the understanding of scaling laws is that such relations inherently assume that there is a flare, an associated CME, and that there will always be a resulting SEP event. However, this one-to-one association scenario is not realistic, since from tens of thousands of recorded flares and CMEs we have only registered a few hundred SEPs \citep{2016JSWSC...6A..42P}. Therefore, scaling laws can offer context under the assumption that solar eruptive events will lead to the acceleration and escape of particles.

The theoretical arguments employed assume that a fraction ($f$) of the magnetic energy stored in an AR is released during a flare, called flare energy  ($E_{flare}$). In the literature this fraction is   estimated to lie in the range 10 to 50\% for large flares \citep{2012ApJ...759...71E, 2012JGRA..117.8103S}. Understandably this fraction has a direct effect on the estimates of the solar--stellar radiation and particle environment \citep[see discussion in][]{herbst2021starspots}.

Furthermore, another point to take into account stems from \cite{1998ApJ...504.1002R} who suggested that energetic particle intensities measured early in a SEP event are bounded by a maximum intensity plateau known as the streaming limit. The mechanism at work is wave generation by particles streaming outward from an intense source near the Sun, and provides a self-regulation of the particle intensity \citep{1994ApJ...424.1032N}. As a result, energetic particles propagating along IMF lines reach a maximum intensity plateau because the scattering processes produced by self-generated waves restrict their streaming. Therefore, according to this scenario, particle intensities measured early in a large SEP event (known as the prompt component of the SEP event) are bounded by a certain upper limit. Scaling laws, in general, do not take into account the streaming limit. However, they provide a simplified quantification of the relation between the driver (i.e., solar eruptive events) and the resulting SEP event, which is useful for the direct approximation of the expected $I_P$ based on $F_{SXR}$ and/or $V_{CME}$.

Extending the scaling relations we provide a direct estimation of the upper limit spectra based on $F_{SXR}$ alone, considering all the caveats and limitations described above.  These spectra are directly usable in the solar case. Admittedly, the form of the spectra deserves a more detailed investigation; nonetheless, the power-law approximation employed in this work provides context since the observed spectral break energies in major SEPs are usually much greater than several tens of MeV and usually fall above 200 MeV,  especially for strong events \citep[see][]{2018ApJ...862...97B}.  

In light of recent results showing that the Sun has been able to give rise to several extreme SEP events, which are likely not manifestations of unknown phenomena, but rather the high-energy--low-probability tail of the ``regular'' SEP distribution \citep{usoskinmind}, we have identified an upper limit to the spectrum of conditions produced by the extreme events:  an upper limit $\sim$X600 SXR flare \citep{2020ApJ...903...41C} based on the AD774/775 cosmogenic nuclide event \citep{miyake2012signature}. We find that the $F_{P}$ at E$>$200 MeV is $\sim$ 10$^{10}$ cm$^{-2}$ and at E$>$430 MeV is $\sim$1.5 $\times$ 10$^{9}$ cm$^{-2}$. In turn, this also means that the Sun produced several extreme solar flares in the past that most likely affected the Earth's radiation environment and evolution. The rationale for the upper limit of X600 for an extreme flare was arrived at in two quite different ways: (1) As noted in Section \ref{sec4}, Cliver et al. (2020b) obtained a SXR class (in the new-scaling) of X400$\pm$200 for the flare inferred for the AD774/775 cosmogenic nuclide event. This value was based on an RMA fit to a  sample of modern GLEs with hard spectra consistent with that deduced for the AD774/775 SEP event (Figure \ref{fig:fig8}). In addition, Cliver et al. (2020b) assumed that the AD774/775 AD flare would be as efficient in producing  high-energy protons as the 1956 flare (in keeping with the use of the 1956 GLE spectrum as the base unit for the AD774/775 AD SEP event). Because of the several month time resolution of the measurements of cosmogenic nuclide concentrations, Cliver et al. (2020b, 2022) also modeled the AD774/775 AD SEP event in terms of multiple equal-contribution eruptions. For example, if the AD774/775 AD event were produced by three such eruptions \citep{cliver2022},  the required peak SXR class  would be reduced to $\sim$X140$\pm$70 (X200$\pm$100 in the new scaling).(2) In contrast to (1), the adjusted to fluence ($F_{P}$) upper limit scaling law (Eq. (3))  that applies the uppermost X600 value from Cliver et al. (2020) in Figure \ref{fig:fig8} was derived from the larger sample of events (listed in Appendix \ref{appendix:C}) used in this work that required observation of E$>$100 MeV protons rather than the E$>$500 MeV protons needed for a GLE.  In Figure \ref{fig:fig8},  the solid orange line spectrum based on a X600 flare for this less energetic sample is consistent (within uncertainties) with the inferred high-energy proton spectrum of the AD774/775 proton event.


The implications of our study extend to other solar-type (see \cite{cliver2022} regarding the distinction between Sun-like and solar-type stars) G-type stars that are known to produce superflares frequently \citep[][]{2012Natur.485..478M, Notsu-etal-2019, Okamoto-etal-2021}. Unless the Sun is a unique star, we can assume that G-type stars show similar behavior. Thus, our understanding of the Sun and its upper limits pertains to the efforts for assessing (within the errors, caveats, and limitations as noted above) their radiation environment and its impact on the habitability of potential exoplanets \citep[see, e.g.,][]{2017ApJ...843...31Y,2019A&A...621A..67H,2019ApJ...874...21F, herbst2021starspots, 2021MNRAS.502.6201B}. In this regard we further included the simulated and modeled  cases of \cite{2022SciA....8I9743H} in our Figure \ref{fig:fig1} (not shown). Their modeled cases fall well within our sample and their obtained scaling for the simulated events are approximately two
orders of magnitude lower than the fluxes estimated by \cite{2016ApJ...833L...8T} and our work. As noted in \cite{2022SciA....8I9743H} the close similarity of the obtained scaling laws' power-law indices suggests that the laws derived from SEP events can be applied to stellar energetic particle events as well. Although there are many more constraints to consider in the stellar case, our work is in agreement with such findings. 

\citet{2016ApJ...833L...8T} applied the concept of scaling laws to assess the upper limit of the Sun on space weather and the terrestrial environment. The current consensus is that the cosmogenic nuclide enhancements in the AD774 were the result of a SEP event \citep{Miyake-etal-2019}. While the detailed observations necessary to explain the outsized SEP event in AD774 are unavailable, it is desirable for worst-case space weather scenarios to make inferences about the circumstances under which it arose, namely  the values of both SXR flare peak intensity and CME speed.  In this paper we investigated the dependence of SEP events on $F_{SXR}$ by applying a range of scaling laws starting with \citet{2016ApJ...833L...8T} to the estimates of flare intensity for the AD774/775 event obtained by \citet{2020ApJ...903...41C, cliver2022}. This study comprises the first of two investigations on the dependence of SEP events on their associated solar activities. In our subsequent work (i.e., part II), we will examine the dependence of SEP events on CMEs.
\begin{acknowledgements}
The authors would like to thank the anonymous referee for the constructive comments that benefited the manuscript at hand. AP, KH, EWC and DL acknowledge the International Space Science Institute and the supported International Team 441: \textit{High EneRgy sOlar partICle Events Analysis (HEROIC)}. KH acknowledges the support of the DFG priority program SPP 1992 “Exploring the Diversity of Extrasolar Planets (HE 8392/1-1)”. AP and KH also acknowledge the supported International Team 464: \textit{The Role Of Solar And Stellar Energetic Particles On (Exo)Planetary Habitability (ETERNAL)}.  Finally, AP and DL acknowledge support from NASA/LWS project NNH19ZDA001N-LWS. DL also acknowledges support from project NNH17ZDH001N-LWS and the Goddard Space Flight Center Heliophysics Innovation Fund (HIF) Program. AMV acknowledges the Austrian Science Fund (FWF): project no. I4555-N.
\end{acknowledgements}

\bibliographystyle{aa}
\bibliography{references}

\appendix

\section{The log($F_{P}$) versus log($I_{p}$) relations} \label{appendix:A}

\begin{figure}[h!]
\centering
\includegraphics[width=0.80\columnwidth]{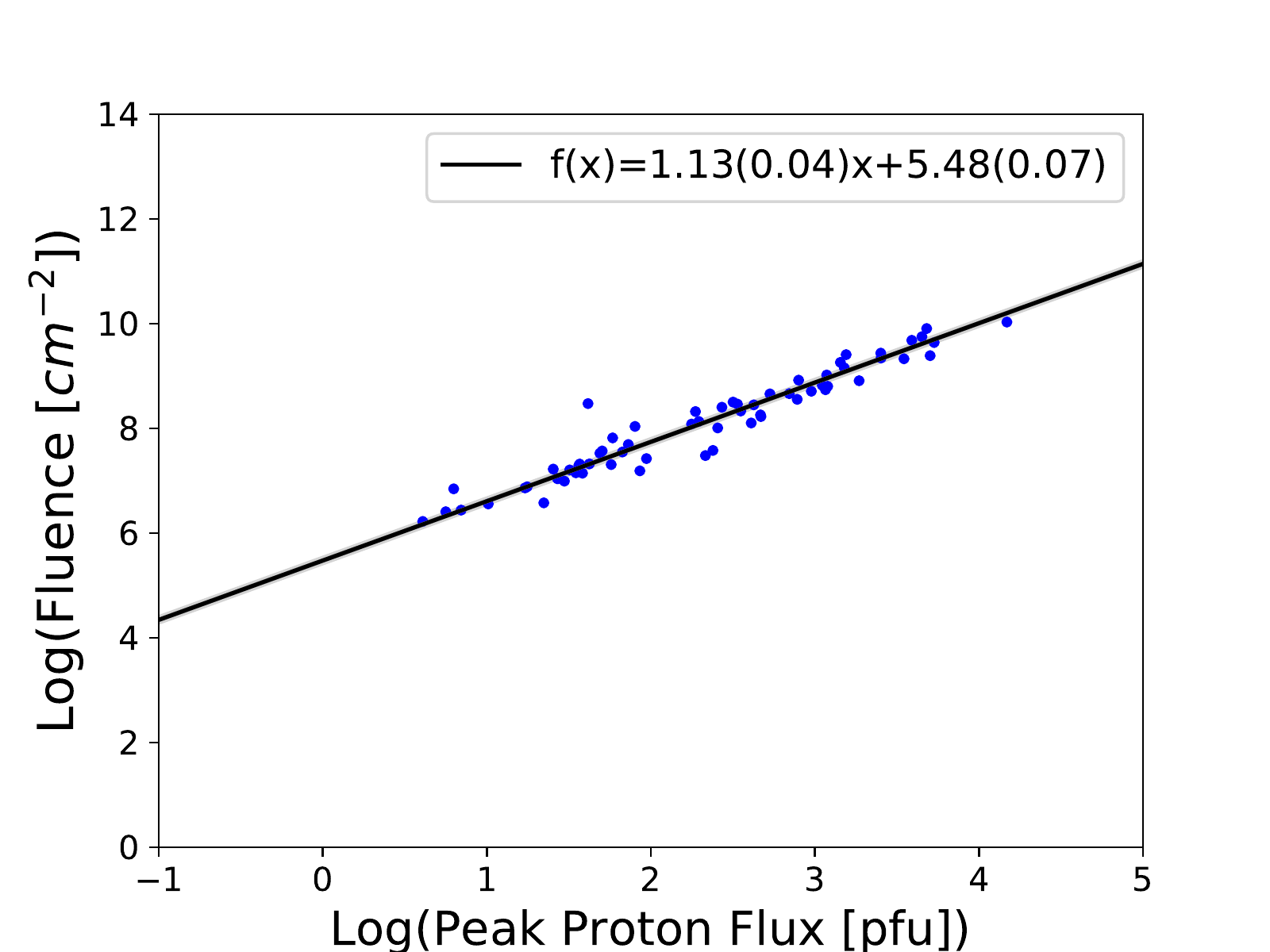}
\includegraphics[width=0.80\columnwidth]{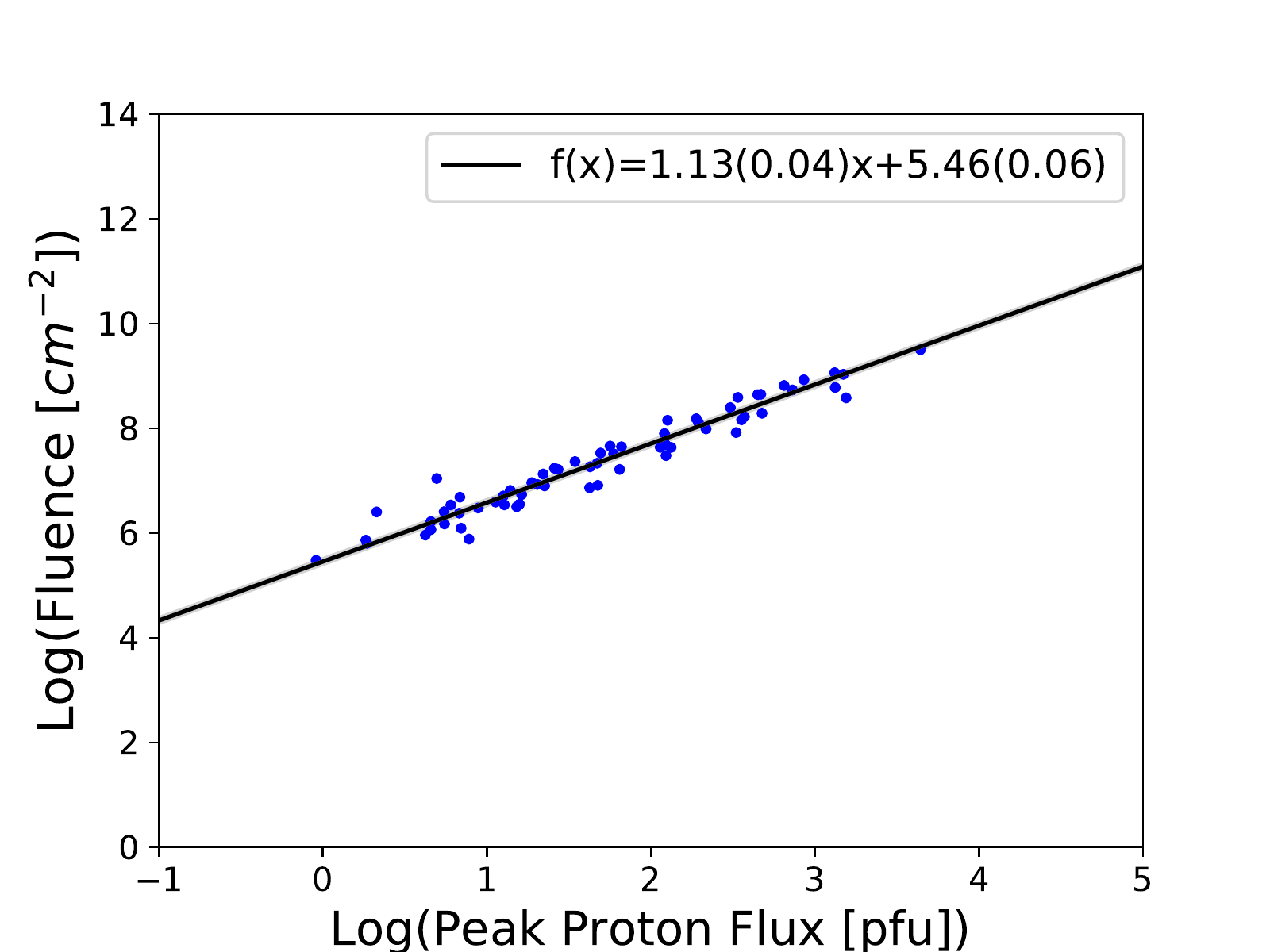}
\includegraphics[width=0.80\columnwidth]{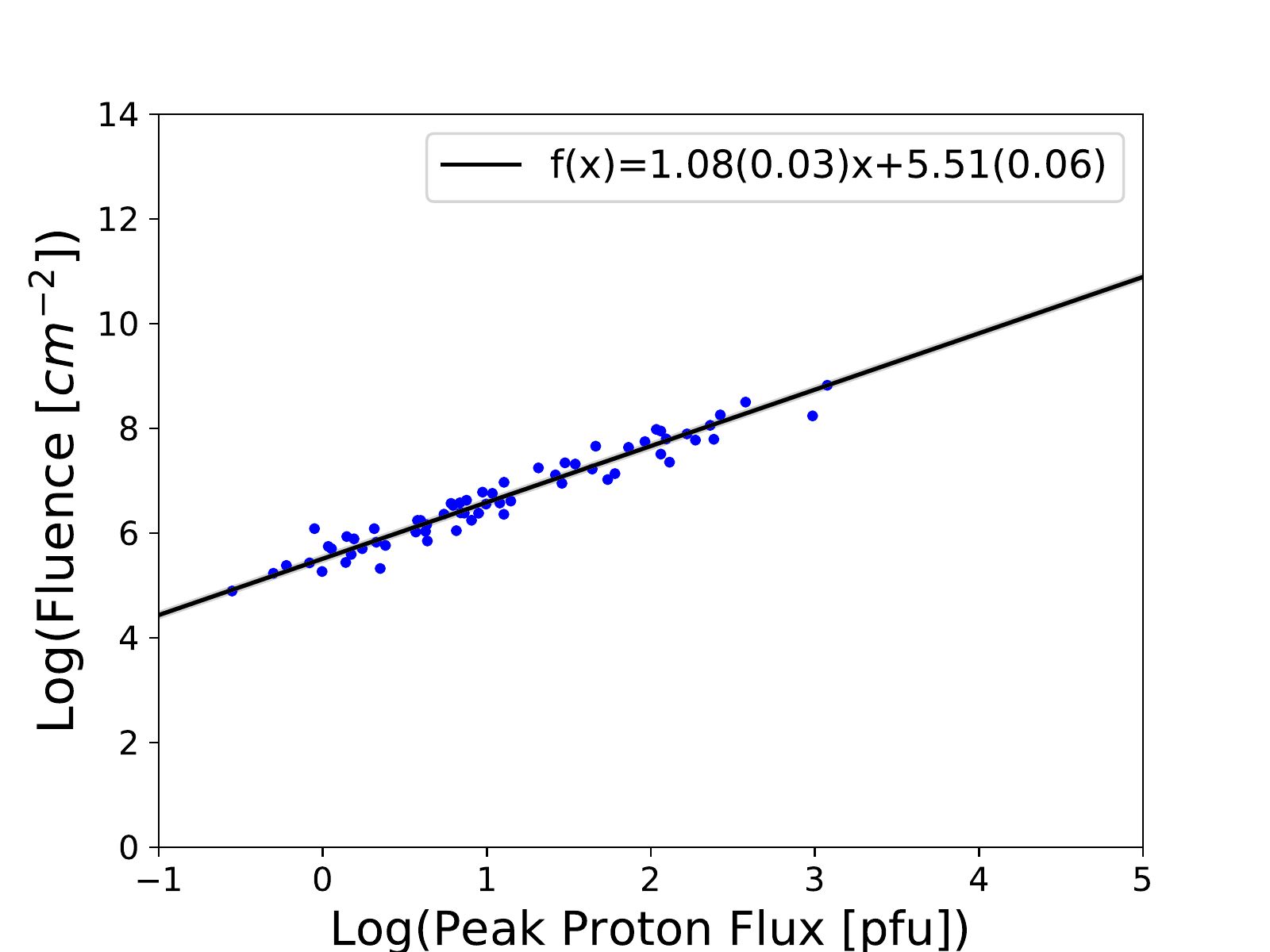}
\includegraphics[width=0.80\columnwidth]{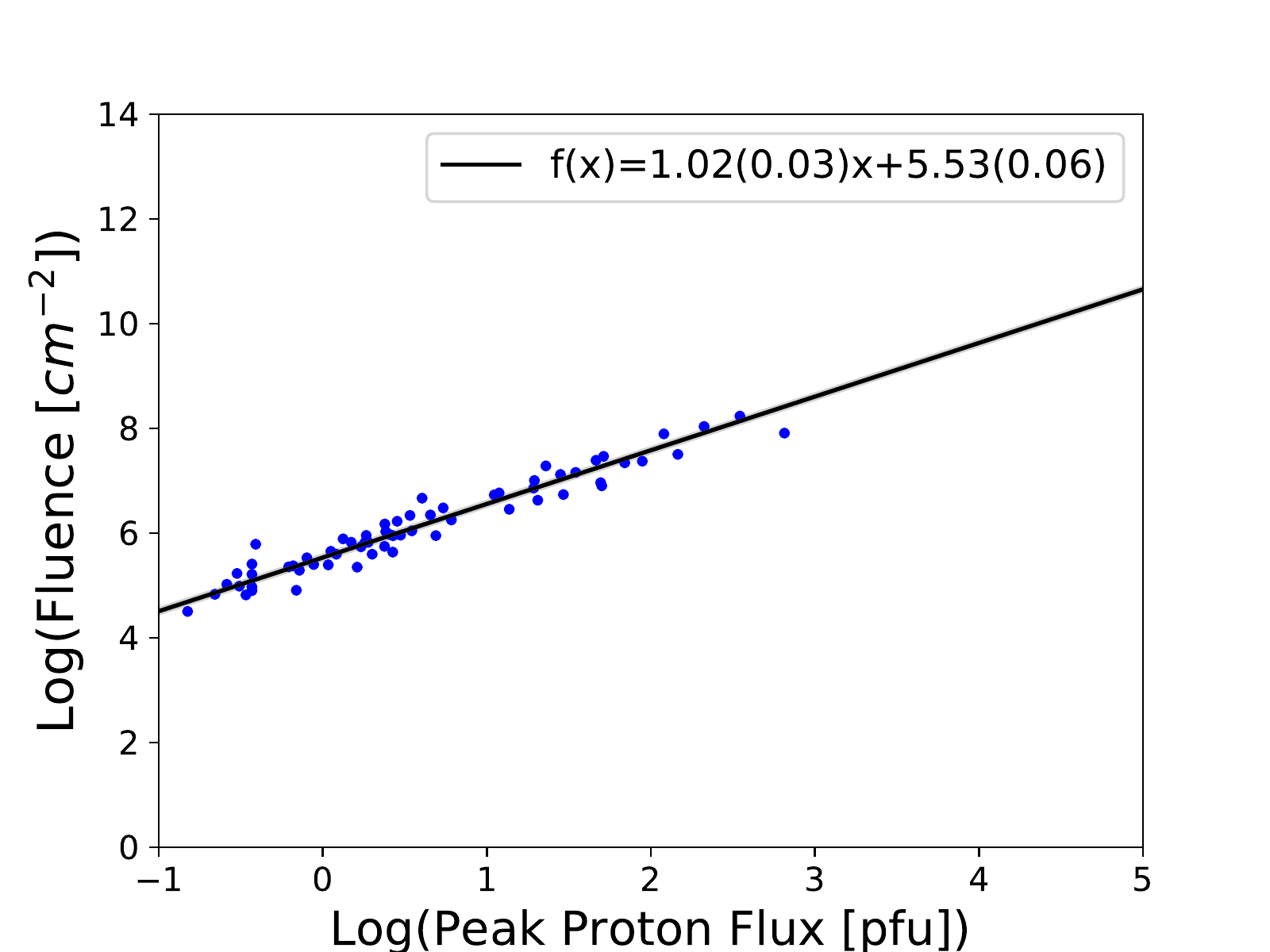}
\caption{Log $F_{P}$ versus log $I_{P}$ for the four integral energy bands of the SEP events:  (from top to bottom) E$>$10-; E$>$30-; E$>$60-; and E$>$100 MeV. In all cases the black solid line represents the RMA regression line.} 
\label{fig:fig5}
\end{figure}

In our examination of the four integral energies employed, we find a very robust statistical linear correlation between logs of $F_{P}$ and $I_{p}$ over an energy range from E$>$10 MeV to E$>$100 MeV and over four decades of measurements (1984--2017), which is the period covered by the 65 SEP events in our sample. Figure \ref{fig:fig5} presents the log-log relation of the $F_{P}$ versus $I_{P}$ for the four integral energy bands of the SEP events  we considered. For all integral energies of interest, we performed linear RMA fits. Table \ref{tab:fluence} summarizes the results. As  can be seen in Figure \ref{fig:fig5}, over approximately four orders of magnitude a robust relation of the $F_{P}$ versus $I_{P}$ is found, as indicated by the high ($\sim$ 0.97) $cc$ values (see Table \ref{tab:fluence}). Thus, over an extensive range of event peak intensities $I_{P}$, these linear fits allow us to make estimates of the event fluences ($F_{P}$) from observed $I_{P}$ within a factor of $\approx$1.65, in agreement with the findings of \cite{kahler2018relating}.

The $F_{P}$ values are vital since they account for both the adiabatic energy losses and multiple traversals of particles across the observer position. The effects of both have been addressed by \cite{chollet2010effects}, who found that the two effects are generally equal and offsetting for a broad range of energies and over many ion species. 


\begin{table}[h!] 
\caption[]{Slope of $F_{P}$-$I_{P}$ and plot amplitude for each integral energy derived in this work.}\label{tab:tabflu}
\label{tab:fluence}
\begin{tabular}{lcccc}
\hline
\bf{Integral} & \bf{Slope $F_{P}$-$I_{P}$} & \bf{Plot amplitude} & \bf{Correlation}\\
 \bf{Energy} & ($\delta$) & & \bf{Coefficient}\\
 \bf{(MeV)}  &                     &  & \bf{(cc)}\\
\hline
E\textgreater{}10   & 1.13$\pm$0.04  & 5.48$\pm$0.07 & 0.96\\
E\textgreater{}30   & 1.13$\pm$0.04  & 5.46$\pm$0.06 & 0.97\\
E\textgreater{}60   & 1.08$\pm$0.03  & 5.51$\pm$0.06 & 0.97\\
E\textgreater{}100  & 1.02$\pm$0.03  & 5.53$\pm$0.06 & 0.97\\
\hline
\end{tabular}%
\end{table}


\begin{table}[h!]
\caption[]{Comparison of fluence values for E$>$100 MeV  derived in this work and in CL16.} \label{tab:tabflucomp1}
\begin{tabular}{lcc}
\hline
              \multicolumn{3}{c}{\textbf{E\textgreater{}100 MeV Fluence  [ $\cdot$ 10$^{3}$ cm$^{-2}$ sr$^{-1}$]}} \\
\hline
\textbf{Date} & \multicolumn{1}{c}{\textbf{this work}} & \multicolumn{1}{c}{\textbf{CL16}} \\
\hline
02/04/2001   & 241   & 220  \\
08/11/2000  & 13300 & 13000\\
17/05/2012  & 338   & 305  \\
24/08/2002  & 434   & 400  \\
20/01/2005  & 6470  & 6400 \\
13/12/2006 & 1880  & 1900 \\
21/04/2002  & 1530  & 1500 \\
26/01/2001  & 637   & 630 \\
\hline
\end{tabular}%
\end{table}

\begin{figure}[h!]
\centering
\includegraphics[width=\columnwidth]{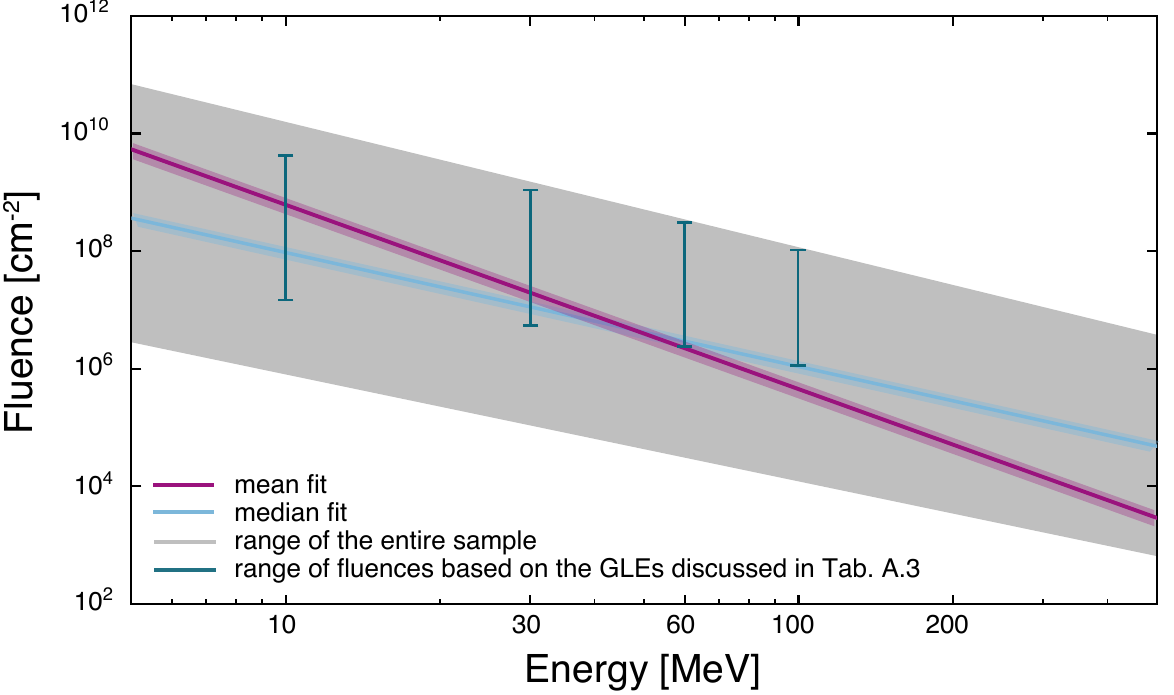}
\caption{Similar to Figure \ref{fig:fig8}. Here the range from the 65 SEP events is presented as a gray shaded area, the magenta line provides the mean, and the blue line represents the median spectrum from the measurements. The blue bars denote the range of the fluence at the respective integral energies from Table \ref{tab:tabflucomp2}.} 
\label{fig:fig14}
\end{figure}

\begin{figure}[h!]
\centering
\includegraphics[width=0.7\columnwidth]{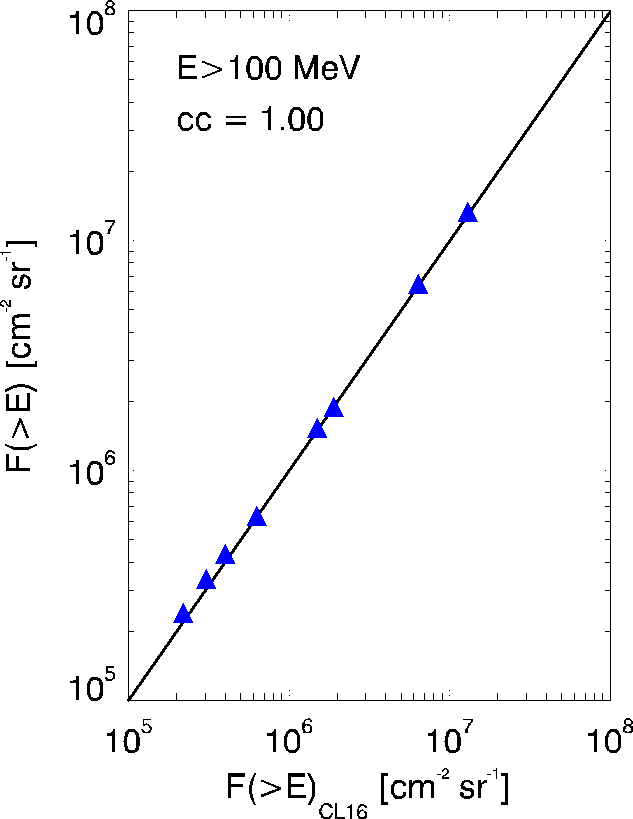}
\caption{Scatter plot of the integral fluences F($>$E) obtained from the data in this work (Y-axis) and from the work of  CL16 (X-axis) for E$>$100 MeV. The blue triangles correspond to individual SEP events (see Table \ref{tab:tabflucomp1}), and the solid black line denotes the diagonal dichotomous.} 
\label{fig:fig12}
\end{figure}

\begin{figure}[h!]
\centering
\includegraphics[width=0.24\textwidth]{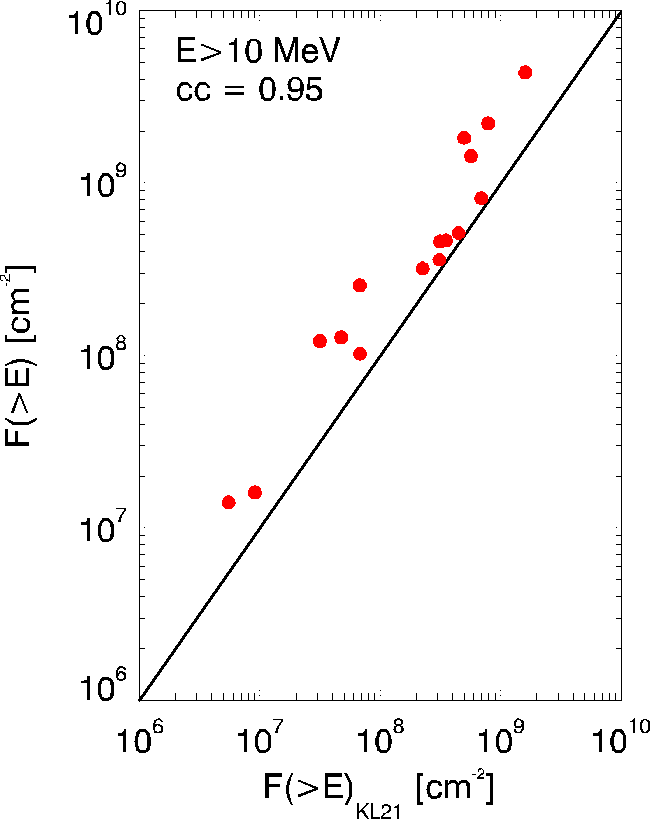}
\includegraphics[width=0.24\textwidth]{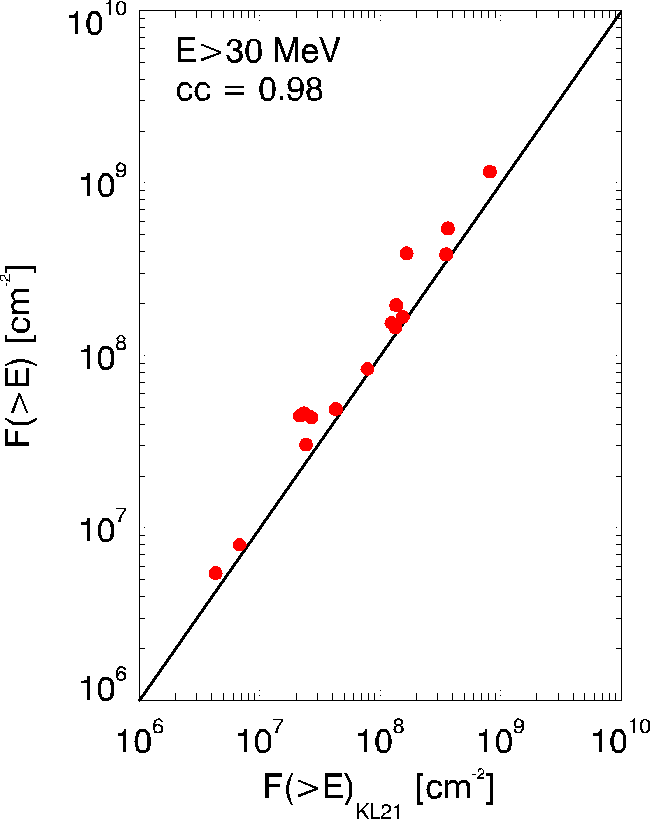}
\includegraphics[width=0.24\textwidth]{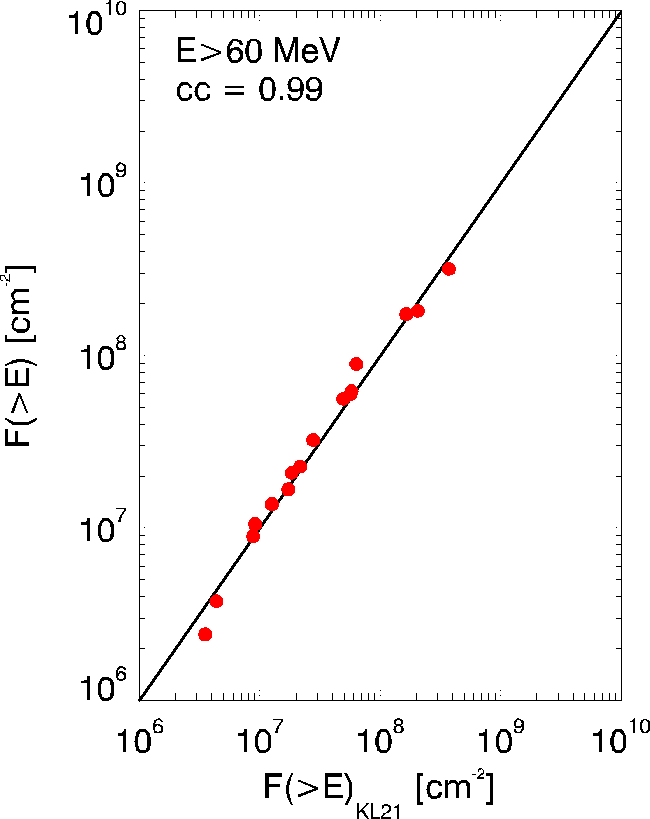}
\includegraphics[width=0.24\textwidth]{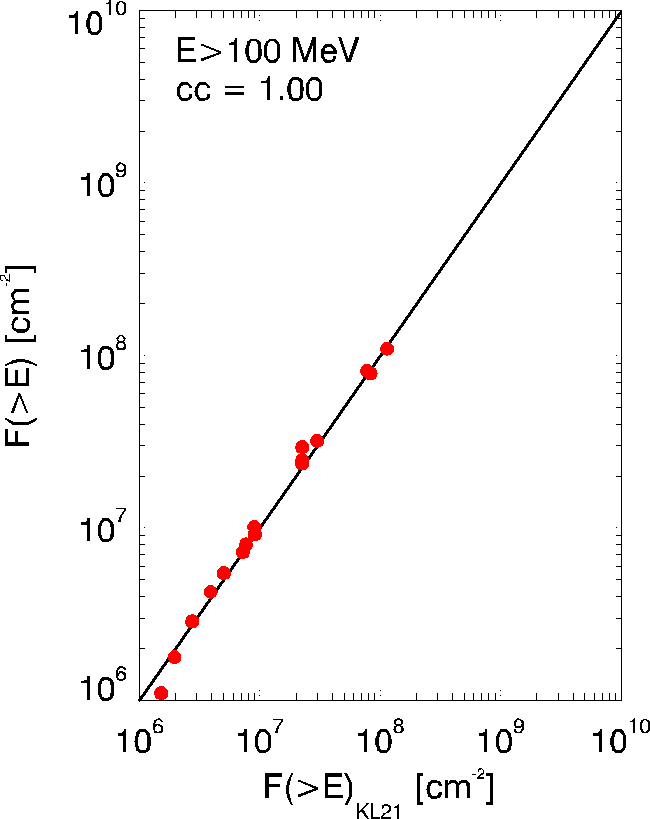}
\caption{Similar to Figure \ref{fig:fig12}, but for scatter plots of the integral fluences F($>$E) obtained from the data in this work (Y-axis) and from the work of  KL21 (X-axis) for each integral energy of interest. The red circles correspond to individual GLE events (see Table \ref{tab:tabflucomp2}), and the solid black line denotes the diagonal dichotomous of each panel.} 
\label{fig:fig13}
\end{figure}

While evaluating the obtained values for the fluence, we surveyed the literature to identify previously published fluences at the respective energies. \citet[][hereafter CL16]{cliver2016flare} lists the fluence at E$>$100 MeV (in protons (cm$^{-2}$ sr$^{-1}$)) for a set of intense SEP events (see their Table 1), while \citet[][hereafter KL21]{koldobskiy2021new} tabulate the fluence  (in protons (cm$^{-2}$)) for all integral energies of interest in this study (E$>$10-; $>$30-; $>$60-; and $>$100 MeV) for 26 GLEs recorded from 1989 to 2017 (see their Table 1). Of these, 16 GLEs were also present in our catalog employed in this study. The comparisons are given in  Tables \ref{tab:tabflucomp1} and \ref{tab:tabflucomp2}. As can be seen, the comparison of the derived fluences shows an excellent agreement between the calculations of this study and the outputs of CL16 (see Table \ref{tab:tabflucomp1}) and Figure \ref{fig:fig12}. The results agree within a mean factor of $\sim$ 1.04. At the same time, the comparison between the obtained omnidirectional fluences in this work and those presented in KL21 show a less strong but still reasonable agreement for E$>$10 MeV (a mean factor of $\sim$ 2.20). However, when shifting to higher energy (i.e., E$>$100 MeV) the agreement gets stronger (a mean factor of $\sim$ 1.02, see Table \ref{tab:tabflucomp2} and Figure \ref{fig:fig13}). In both comparisons, a close relationship between the obtained fluences at each energy obtained in this work and CL16 and KL21 is evident by the high cc ($\ge$0.95). However, it  should be noted,  as highlighted in Section \ref{sec:data}, that different GOES spacecraft lead to differences between the intensities. As a result, this affects the derived peak proton flux and consequently the calculated fluence at each case. For example,  for the 16 GLEs, KL21 and this study used measurements from the same GOES spacecraft for only 7 events ($\sim$43\% of the cases). If the comparison considers only these seven events, the mean factor for E$>$10 MeV falls to $\sim$1.30. Figure \ref{fig:fig14} is similar to Figure \ref{fig:fig8}. It presents the fluence range from the 65 SEP events with a gray shaded area and the mean(median) spectrum obtained from the measurements as blue(magenta) lines. The blue bars denote the range of the fluence at the respective integral energies from Table \ref{tab:tabflucomp2}. As a result, the obtained fluence values for the GLEs seem to lie within the gray shaded area.       

\begin{table}[h!]
\caption[]{Comparison of fluence values for E$>$10-; E$>$30-; E$>$60; and E$>$100 MeV derived in this work and in KL21.} \label{tab:tabflucomp2}
\begin{tabular}{lcccc}
\hline
\multicolumn{5}{c}{\textbf{Fluence  [cm$^{-2}$]}}\\
\hline
              & \multicolumn{2}{r}{\textbf{E\textgreater{}10 MeV }} 
              & \multicolumn{2}{r}{\textbf{E\textgreater{}30 MeV }}\\
\hline
\textbf{Event} & \multicolumn{1}{c}{\textbf{this work}} & \multicolumn{1}{c}{\textbf{KL21}} & \multicolumn{1}{c}{\textbf{this work}} & \multicolumn{1}{c}{\textbf{KL21}} \\
\hline
GLE40   &       1.60E+07        &       9.18E+06        &       7.97E+06        &       6.84E+06        \\
GLE41   &       1.82E+09        &       5.02E+08        &       3.90E+08        &       1.67E+08        \\
GLE44   &       4.36E+09        &       1.62E+09        &       1.16E+09        &       8.20E+08        \\
GLE45   &       2.21E+09        &       7.94E+08        &       5.43E+08        &       3.68E+08        \\
GLE46   &       1.40E+07        &       5.54E+06        &       5.45E+06        &       4.33E+06        \\
GLE47   &       1.27E+08        &       4.78E+07        &       4.35E+07        &       2.71E+07        \\
GLE48   &       1.21E+08        &       3.18E+07        &       4.60E+07        &       2.35E+07        \\
GLE52   &       2.54E+08        &       6.81E+07        &       4.46E+07        &       2.17E+07        \\
GLE55   &       4.56E+08        &       3.17E+08        &       1.54E+08        &       1.25E+08        \\
GLE60   &       5.12E+08        &       4.51E+08        &       1.45E+08        &       1.35E+08        \\
GLE63   &       3.58E+08        &       3.13E+08        &       8.34E+07        &       7.91E+07        \\
GLE64   &       3.18E+08        &       2.27E+08        &       4.87E+07        &       4.31E+07        \\
GLE67   &       1.43E+09        &       5.73E+08        &       1.95E+08        &       1.37E+08        \\
GLE69   &       8.13E+08        &       6.98E+08        &       3.84E+08        &       3.57E+08        \\
GLE70   &       4.62E+08        &       3.55E+08        &       1.67E+08        &       1.55E+08        \\
GLE71   &       1.02E+08        &       6.85E+07        &       3.03E+07        &       2.44E+07        \\
\hline
              & \multicolumn{2}{r}{\textbf{E\textgreater{}60 MeV }} &\multicolumn{2}{r}{\textbf{E\textgreater{}100 MeV }}\\
\hline
\textbf{Event} & \multicolumn{1}{c}{\textbf{this work}} & \multicolumn{1}{c}{\textbf{KL21}} & \multicolumn{1}{c}{\textbf{this work}} & \multicolumn{1}{c}{\textbf{KL21}} \\
\hline
GLE40   &       3.75E+06        &       4.40E+06        &       1.78E+06        &       1.97E+06        \\
GLE41   &       8.91E+07        &       6.39E+07        &       2.93E+07        &       2.27E+07        \\
GLE44   &       3.17E+08        &       3.74E+08        &       1.09E+08        &       1.15E+08        \\
GLE45   &       1.81E+08        &       2.07E+08        &       7.85E+07        &       8.39E+07        \\
GLE46   &       2.41E+06        &       3.55E+06        &       1.10E+06        &       1.53E+06        \\
GLE47   &       1.67E+07        &       1.74E+07        &       7.24E+06        &       7.28E+06        \\
GLE48   &       2.08E+07        &       1.85E+07        &       1.01E+07        &       9.08E+06        \\
GLE52   &       8.93E+06        &       8.85E+06        &       2.86E+06        &       2.77E+06        \\
GLE55   &       5.59E+07        &       4.97E+07        &       2.46E+07        &       2.27E+07        \\
GLE60   &       6.19E+07        &       5.83E+07        &       3.19E+07        &       3.02E+07        \\
GLE63   &       2.26E+07        &       2.18E+07        &       8.00E+06        &       7.74E+06        \\
GLE64   &       1.37E+07        &       1.27E+07        &       5.45E+06        &       5.06E+06        \\
GLE67   &       3.23E+07        &       2.80E+07        &       9.18E+06        &       9.19E+06        \\
GLE69   &       1.73E+08        &       1.66E+08        &       8.13E+07        &       7.85E+07        \\
GLE70   &       5.96E+07        &       5.70E+07        &       2.36E+07        &       2.27E+07        \\
GLE71   &       1.05E+07        &       9.21E+06        &       4.25E+06        &       3.93E+06        \\
\hline
\end{tabular}%
\end{table}

\begin{figure}[h!]
\centering
\includegraphics[width=0.48\columnwidth]{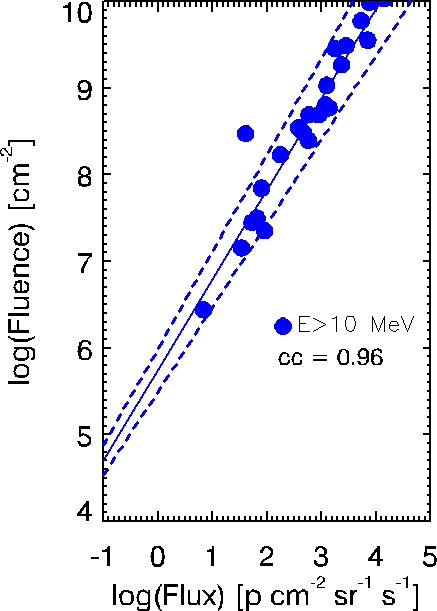}
\includegraphics[width=0.48\columnwidth]{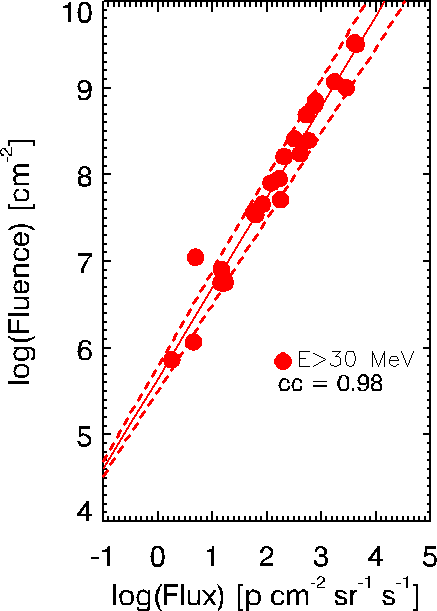}
\includegraphics[width=0.48\columnwidth]{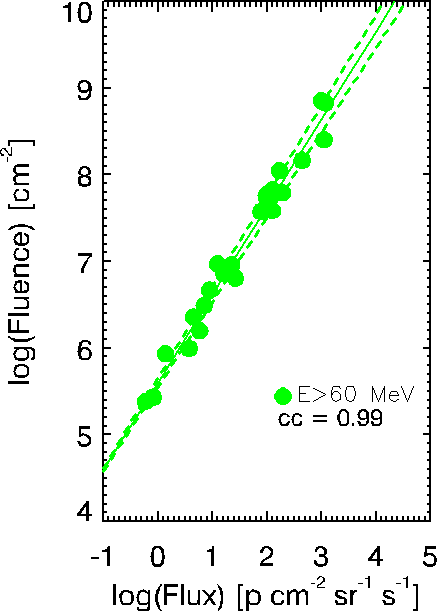}
\includegraphics[width=0.48\columnwidth]{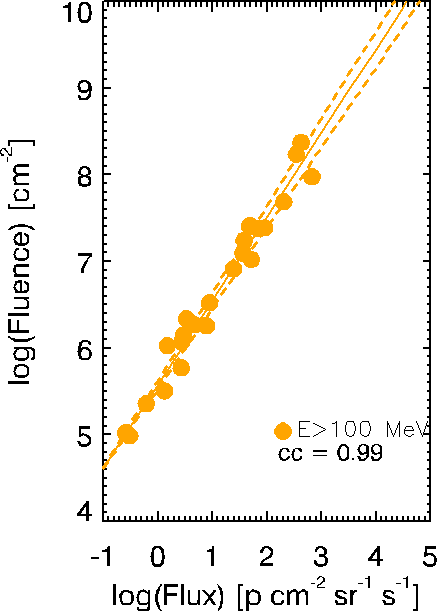}
\caption{Similar to Figure \ref{fig:fig5}, but for annual values. Each integral energy is presented with a different color (E$>$10 MeV - blue; E$>$30 MeV - red; E$>$60 MeV - green; and E$>$100 MeV - orange). The obtained linear fit is depicted as a solid line in each panel, while the dotted lines give the 1$\sigma$ error of the fit. The corresponding correlation coefficients ($cc$) are also presented in each plot.} 
\label{fig:fig5_ann}
\end{figure}

As a next step, the annual fluence versus the annual peak proton flux for each integral energy of interest was investigated. As stated before, we find a very robust statistical linear correlation between logs of $F_{P}$ and $I_{p}$ for each integral energy. Therefore, the linear relations seem to be evident in the annual case as well. Figure \ref{fig:fig5_ann} presents the log-log relation of  $F_{P}$ versus $I_{P}$ for the four integral energy bands. For all integral energies of interest, we performed linear RMA fits. As  can be seen in Figure \ref{fig:fig5_ann} over approximately four orders of magnitude a robust relation of  $F_{P}$ versus $I_{P}$ is found. Again, in the annual case high $cc$ values ($\sim$ 0.97)
are obtained (see Table \ref{tab:tabflu_ann}).

\begin{table}[h!] 
\caption[]{Slope of $F_{P}$-$I_{P}$ and plot amplitude for each integral energy for the annual case derived in this work.}\label{tab:tabflu_ann}
\label{tab:fluence_ann}
\begin{tabular}{lcccc}
\hline
\bf{Integral} & \bf{Slope $F_{P}$-$I_{P}$} & \bf{Plot amplitude} & \bf{Correlation}\\
 \bf{Energy} & (annual) & & \bf{Coefficient}\\
 \bf{(MeV)}  &                     &  & \bf{(cc)}\\
\hline
E\textgreater{}10   & 1.09$\pm$0.04  & 5.62$\pm$0.18 & 0.96\\
E\textgreater{}30   & 1.07$\pm$0.04  & 5.61$\pm$0.12 & 0.98\\
E\textgreater{}60   & 1.02$\pm$0.03  & 5.59$\pm$0.05 & 0.99\\
E\textgreater{}100  & 0.98$\pm$0.04  & 5.56$\pm$0.05 & 0.99\\
\hline
\end{tabular}%
\end{table}


\clearpage

\section{The 8 November 2000 event} \label{appendix:B}
\begin{figure}[h!]
\centering
\includegraphics[width=0.48\columnwidth]{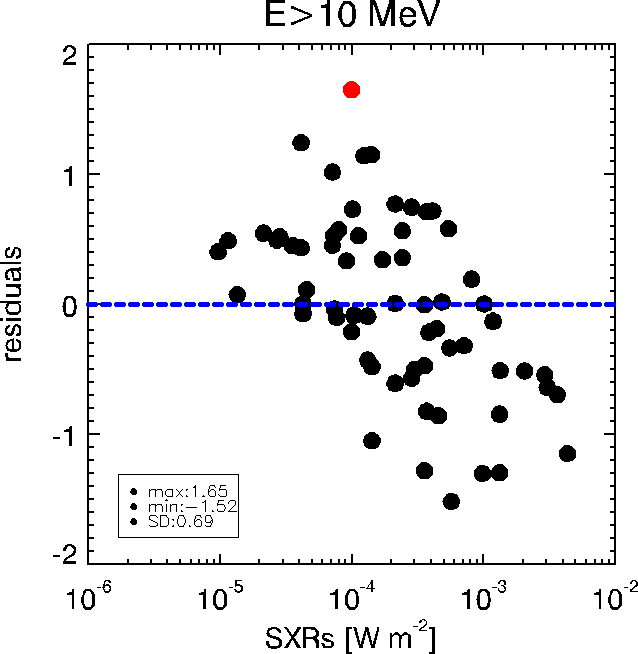}
    \vspace{0.5cm}
\includegraphics[width=0.48\columnwidth]{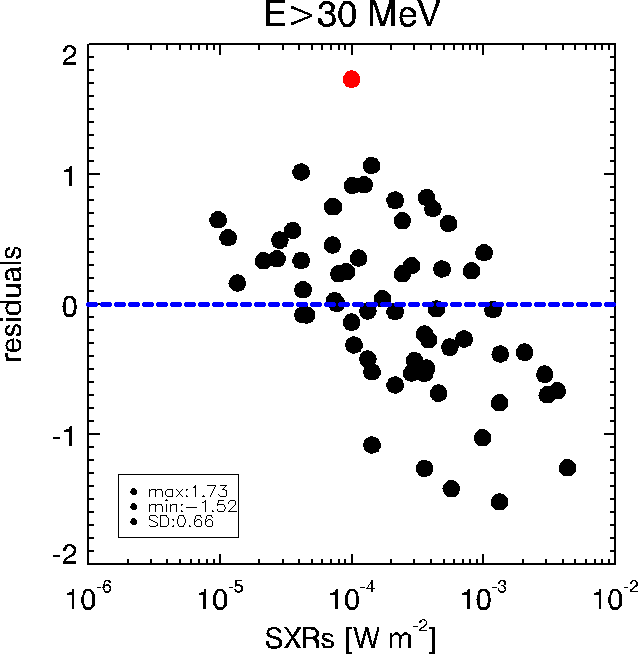}
\includegraphics[width=0.48\columnwidth]{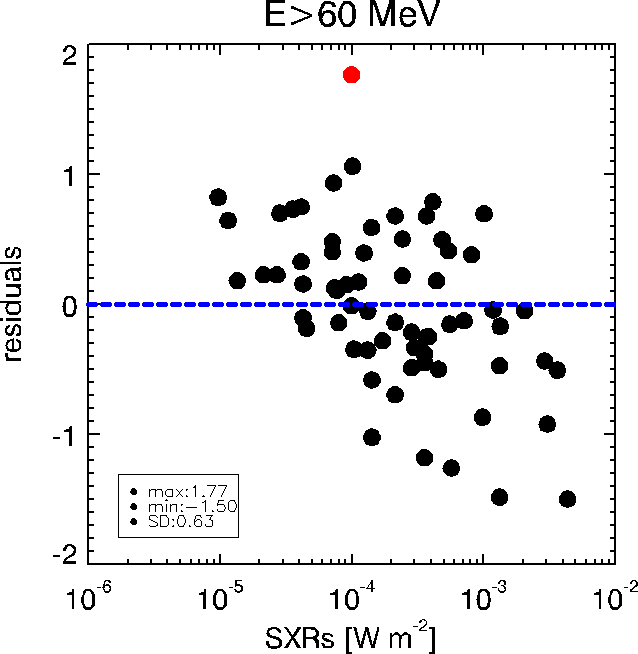}
\includegraphics[width=0.48\columnwidth]{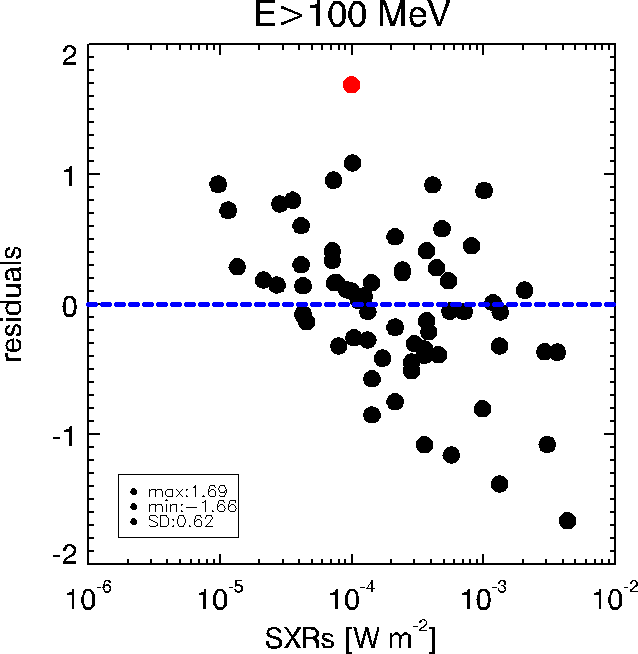}
\caption{Calculated  residuals vs. the SXR flux for all 65 events in our sample. Each panel corresponds to an integral energy (E$>$10-; E$>$30-; E$>$60-; E$>$100 MeV). In each panel the red point depicts the 8 November 2000 event. The legend provides the standard deviation  of the residuals, as well as the minimum and the maximum residual of the sample.} 
\label{fig:residuals}
\end{figure}

The 8 November 2000 event is  clearly distinguished as the one with the largest obtained peak proton flux for a relatively modest SXR flare. We assess the departure of this particular event from the linear fit obtained in Figures \ref{fig:fig1} and \ref{fig:fig4} by calculating the residuals as $residual = predicted~y - actual~y $, and the S.D.  of the residuals. Figure \ref{fig:residuals} demonstrates these calculated residuals for each integral energy of interest (i.e., E$>$10; E$>30$; E$>$60 and E$>$100 MeV) as black circles. The residuals lie on the vertical axis, while the SXR magnitude (i.e., the independent variable) is on the horizontal axis. The dotted blue line represents the perfect agreement of the fitted value to the actual one. The $S.D.$ for each integral energy is presented on each panel of Figure \ref{fig:residuals}. It ranges from $\rm  S.D.$=0.68 (E$>$10 MeV) to $\rm  S.D.$=0.62 (E$>$100 MeV). The 8 November 2000 event is indicated with a red circle. As it can be seen, there is a random dispersion of all points around the perfect fit (i.e., the blue dotted line). Moreover, all points seem to have a large spread with the minimum and maximum residual printed in the legend of each panel of Figure \ref{fig:residuals}. Finally, the residuals of 8 November 2000 event lie well within 3-$\rm S.D.$ in each integral energy of interest.

\clearpage

\section{Complete list of SEP events utilized in this study}
Here we provide a listing of the SEP events, their achieved peak proton flux, and fluence at each integral energy of interest, as well as their parent solar events. \label{appendix:C}

\begin{table*}[]
\caption{The 65 SEP events (1984-2017) that extend to E$>$100 MeV. For each event the calculated fluence (cm$^{-2}$) and the peak proton flux (pfu | cm$^{-2}$ s$^{-1}$ sr$^{-1}$ MeV$^{-1}$) for each integral energy is given. The GOES satellite used in the identification of the event is further tabulated. Solar associations include the characteristics of the associated solar flares and CMEs.}
\resizebox{0.90\textwidth}{!}{%
\begin{tabular}{llllllllllllllllllll}
   & \multicolumn{11}{c}{\textbf{Proton Event}}                                                                                                                                                                                                     &           & \multicolumn{5}{c}{\textbf{Solar Flare Event}}                                 & \multicolumn{2}{c}{\textbf{Coronal Mass Ejection (CME)}} \\
   \hline
   & \multicolumn{3}{l}{\textbf{Start time of the event}} & \multicolumn{4}{c}{\textbf{Fluence {[]cm$^{-2}$]}}}                                                & \multicolumn{4}{c}{\textbf{Peak Proton Flux [pfu]}}                                             & \textbf{GOES}      & \multicolumn{2}{l}{\textbf{Start Time}} & \multicolumn{2}{c}{\textbf{Position}} & \textbf{X-ray}  & \textbf{Width} & \textbf{Speed}  \\
   \hline
\textbf{No} & \textbf{Year}      & \textbf{Date}           & \textbf{Time (UT)}      & \textbf{\textgreater{}10 MeV} & \textbf{\textgreater{}30 MeV} & \textbf{\textgreater{}60 MeV} & \textbf{\textgreater{}100 MeV} & \textbf{\textgreater{}10 MeV} & \textbf{\textgreater{}30 MeV} & \textbf{\textgreater{}60 MeV} & \textbf{\textgreater{}100 MeV} & \textbf{satellite} & \textbf{Date}         & \textbf{Time (UT)}       & \textbf{Lon}           & \textbf{Lat}          & \textbf{Peak}   & \textbf{(deg)} & \textbf{(Km/s)} \\
\hline
1 & 1984 & 14-Mar & 4:05 & 3.14E+07 & 8.05E+06 & 3.09E+06 & 1.41E+06 & 66.61 & 14.95 & 7.19 & 3.01 &  & 14-Mar & 3:15 & 42 & -12 & M2 &  &    \\
2 & 1985 & 22-Jan & 1:20 & 7.01E+06 & 2.54E+06 & 1.22E+06 & 6.13E+05 & 6.28 & 2.13 & 0.89 & 0.39 &  & 21-Jan & 23:52 & 38 & -8 & X4 &  &   \\
3 & 1985 & 9-Jul & 2:25 & 1.54E+07 & 3.19E+06 & 1.05E+06 & 4.50E+05 & 85.68 & 15.21 & 3.69 & 1.12 &  & 9-Jul & 1:26 & 27 & -13 & M2.9 &  &  \\
4 & 1986 & 7-Feb & 13:00 & 1.36E+08 & 2.16E+07 & 3.58E+06 & 8.90E+05 & 196 & 47.1 & 9.89 & 2.68 & 5 & 7-Feb & 10:11 & 20 & -10 & M5 &  &    \\
5 & 1986 & 10-Feb & 21:00 & 2.57E+06 & 6.39E+05 & 1.71E+05 & 6.77E+04 & 5.62 & 1.86 & 0.5 & 0.22 & 5 & 10-Feb & 20:25 & 32 & -1 & C9.5 &   &  \\
6 & 1986 & 14-Feb & 10:35 & 2.10E+08 & 2.32E+07 & 3.39E+06 & 9.03E+05 & 187 & 34.5 & 6.22 & 1.84 & 5 & 14-Feb & 9:10 & 76 & 1 & M6.4 &  &   \\
7 & 1988 & 12-Oct & 3:40 & 2.75E+06 & 7.33E+05 & 2.41E+05 & 1.05E+05 & 6.97 & 1.83 & 0.6 & 0.26 & 7 & 12-Oct & 4:57 & 66 & -20 & X2.5 &  &  \\
8 & 1989 & 23-Mar & 20:30 & 9.82E+06 & 1.24E+06 & 2.74E+05 & 7.98E+04 & 29.7 & 6.97 & 1.38 & 0.37 & 7 & 23-Mar & 19:25 & 28 & 18 & X1.5 &   &  \\
9 & 1989 & 18-Jun & 15:00 & 3.60E+06 & 1.50E+06 & 5.82E+05 & 2.47E+05 & 10.2 & 5.52 & 2.41 & 1.08 & 7 & 18-Jun & 16:19 & 57 & 15 & C6.8 &   &  \\
10 & 1989 & 25-Jul & 9:05 & 1.60E+07 & 7.96E+06 & 3.75E+06 & 1.78E+06 & 32 & 22.5 & 12 & 6.08 & 7 & 25-Jul & 8:39 & 84 & 25 & X2.6 &  &    \\
11 & 1989 & 12-Aug & 15:30 & 5.64E+09 & 1.08E+09 & 1.14E+08 & 1.44E+07 & 4490 & 1490 & 230 & 34.8 & 7 & 12-Aug & 13:57 & 37 & -16 & X2.6 &  &  \\
12 & 1989 & 16-Aug & 1:30 & 1.82E+09 & 3.90E+08 & 8.90E+07 & 2.93E+07 & 1430 & 339 & 115 & 51.5 & 7 & 16-Aug & 1:08 & 84 & -18 & X20 &  &  \\
13 & 1989 & 22-Oct & 16:30 & 4.36E+09 & 1.16E+09 & 3.17E+08 & 1.09E+08 & 5330 & 1320 & 378 & 211 & 7 & 22-Oct & 17:08 & 31 & -27 & X2.9 &  &  \\
14 & 1989 & 24-Oct & 19:00 & 2.21E+09 & 5.43E+08 & 1.81E+08 & 7.85E+07 & 2530 & 729 & 265 & 120 & 7 & 24-Oct & 17:36 & 57 & -30 & X5.7 &   &  \\
15 & 1989 & 15-Nov & 7:05 & 1.40E+07 & 5.44E+06 & 2.41E+06 & 1.10E+06 & 38.3 & 16.3 & 7.28 & 3.49 & 7 & 15-Nov & 6:38 & 26 & 11 & X3.2 &   &  \\
16 & 1989 & 30-Nov & 13:15 & 2.14E+09 & 1.29E+08 & 6.05E+06 & 3.94E+05 & 3490 & 195 & 9.42 & 1.21 & 7 & 30-Nov & 11:45 & 52 & 24 & X2 &   &  \\
17 & 1990 & 21-May & 22:25 & 1.27E+08 & 4.35E+07 & 1.67E+07 & 7.24E+06 & 409 & 114 & 43.9 & 19.3 & 7 & 21-May & 22:12 & 36 & 35 & X5 &  &  \\
18 & 1990 & 24-May & 20:35 & 1.21E+08 & 4.60E+07 & 2.08E+07 & 1.01E+07 & 177 & 56.4 & 34.6 & 19.5 & 7 & 24-May & 20:46 & 78 & 33 & X9.3 &  &  \\
19 & 1991 & 15-Jun & 8:50 & 1.05E+09 & 2.50E+08 & 6.30E+07 & 2.20E+07 & 1180 & 305 & 124 & 69.3 & 7 & 15-Jun & 6:33 & 69 & 33 & X12 &  &   \\
20 & 1991 & 30-Oct & 6:55 & 2.65E+07 & 9.22E+06 & 3.70E+06 & 1.68E+06 & 94 & 18.8 & 6.05 & 2.84 & 7 & 30-Oct & 6:11 & 25 & -8 & X2.5 &  &  \\
21 & 1992 & 25-Jun & 20:15 & 2.54E+08 & 4.46E+07 & 8.92E+06 & 2.86E+06 & 271 & 66.2 & 28.7 & 13.7 & 7 & 25-Jun & 19:47 & 67 & 9 & X3.9 &  &  \\
22 & 1992 & 30-Oct & 18:25 & 2.56E+09 & 4.48E+08 & 4.33E+07 & 5.38E+06 & 1550 & 468 & 73.1 & 11.1 & 7 & 30-Oct & 17:02 & 61 & -22 & X1.7  &  &  \\
23 & 1993 & 4-Mar & 12:40 & 7.30E+06 & 1.66E+06 & 5.03E+05 & 1.94E+05 & 17.1 & 4.56 & 1.74 & 0.72 & 7 & 4-Mar & 12:17 & 56 & -14 & C8.1 &  &  \\
24 & 1993 & 12-Mar & 18:30 & 2.08E+07 & 3.91E+06 & 1.08E+06 & 3.96E+05 & 36.8 & 11.3 & 4.23 & 2 & 7 & 12-Mar & 16:48 & 51 & 0 & M7 &  &   \\
25 & 1994 & 19-Oct & 21:45 & 1.41E+07 & 1.17E+06 & 2.70E+05 & 9.69E+04 & 34.9 & 4.56 & 0.83 & 0.31 & 7 & 19-Oct & 22:35 & 24 & 12 & M3.2 &  &  \\
26 & 1997 & 4-Nov & 6:05 & 3.54E+07 & 8.49E+06 & 2.44E+06 & 9.43E+05 & 67.1 & 20.3 & 6.93 & 2.55 & 8 & 4-Nov & 5:52 & 33 & -14 & X2.1 & 360 & 785\\
27 & 1997 & 6-Nov & 12:20 & 4.56E+08 & 1.54E+08 & 5.59E+07 & 2.46E+07 & 532 & 189 & 92 & 46.3 & 8 & 6-Nov & 11:49 & 63 & -18 & X9.4 & 360 & 1556\\
28 & 1998 & 6-May & 8:15 & 3.79E+07 & 8.18E+06 & 2.28E+06 & 8.97E+05 & 239 & 47.5 & 12.7 & 4.89 & 8 & 6-May & 7:58 & 65 & -11 & X2.7 & 248 & 1099 \\
29 & 1998 & 30-Sep & 14:10 & 5.48E+08 & 4.33E+07 & 4.09E+06 & 9.12E+05 & 1160 & 133 & 14 & 2.98 & 8 & 30-Sep & 13:04 & 85 & 19 & M2.9 & ..... & .....\\
30 & 2000 & 10-Jun & 17:00 & 2.09E+07 & 3.47E+06 & 7.07E+05 & 2.23E+05 & 42.2 & 12.8 & 4.34 & 1.62 & 8 & 10-Jun & 16:40 & 38 & 22 & M5.2 & 360 & 1108\\
31 & 2000 & 22-Jul & 11:50 & 7.66E+06 & 9.17E+05 & 1.85E+05 & 6.59E+04 & 17.6 & 4.22 & 0.99 & 0.34 & 8 & 22-Jul & 11:17 & 56 & 14 & M3 & 259 & 1230\\
32 & 2000 & 8-Nov & 23:20 & 1.07E+10 & 3.18E+09 & 6.67E+08 & 1.73E+08 & 14800 & 4400 & 1190 & 349 & 8 & 8-Nov & 22:42 & 77 & 10 & M7 & 170 & 1738\\
33 & 2001 & 28-Jan & 16:45 & 3.36E+07 & 3.45E+06 & 5.56E+05 & 1.69E+05 & 48.9 & 6.03 & 1.08 & 0.3 & 8 & 28-Jan & 15:40 & 59 & -4 & M1.5 & 360 & 916\\
34 & 2001 & 2-Apr & 11:20 & 1.66E+06 & 3.03E+05 & 7.82E+04 & 3.19E+04 & 4.07 & 0.91 & 0.28 & 0.15 & 8 & 2-Apr & 10:58 & 62 & 17 & X1 & 80 & 992 \\
35 & 2001 & 2-Apr & 23:15 & 6.61E+08 & 9.79E+07 & 1.29E+07 & 3.02E+06 & 1110 & 217 & 26.2 & 5.42 & 8 & 2-Apr & 21:32 & 82 & 14 & X20 & 244 & 2505 \\
36 & 2001 & 12-Apr & 11:20 & 3.71E+07 & 6.54E+06 & 1.74E+06 & 6.69E+05 & 50.5 & 13.9 & 3.95 & 1.49 & 8 & 12-Apr & 9:39 & 43 & -19 & X2 & 360 & 1184\\
37 & 2001 & 15-Apr & 13:50 & 5.12E+08 & 1.45E+08 & 6.19E+07 & 3.19E+07 & 951 & 357 & 242 & 146 & 8 & 15-Apr & 13:19 & 85 & -20 & X14.4 & 167 & 1199\\
38 & 2001 & 23-Nov & 1:05 & 8.08E+09 & 8.47E+08 & 4.57E+07 & 4.65E+06 & 4800 & 857 & 46.1 & 4.03 & 8 & 23.Nov & 22:38 & 36 & -17 & M9.9 & 360 & 1437\\
39 & 2001 & 26-Dec & 5:45 & 3.58E+08 & 8.33E+07 & 2.26E+07 & 8.00E+06 & 780 & 331 & 130 & 50.2 & 8 & 26-Dec & 4:32 & 54 & 8 & M7.1 & 212 & 1446\\
40 & 2002 & 21-Apr & 1:40 & 2.73E+09 & 6.59E+08 & 9.56E+07 & 1.92E+07 & 2520 & 649 & 108 & 22.9 & 8 & 21-Apr & 0:43 & 84 & -14 & X1.5 & 360 & 2393 \\
41 & 2002 & 22-Aug & 2:25 & 1.97E+07 & 5.12E+06 & 1.45E+06 & 5.47E+05 & 36.4 & 12.6 & 4.3 & 1.71 & 8 & 22-Aug & 1:47 & 62 & -7 & M5.4 & 360 & 998 \\
42 & 2002 & 24-Aug & 1:15 & 3.17E+08 & 4.87E+07 & 1.37E+07 & 5.45E+06 & 317 & 123 & 60.4 & 29.3 & 8 & 24-Aug & 0:49 & 81 & -2 & X3.1 & 360 & 1913\\
43 & 2003 & 31-May & 2:40 & 1.09E+07 & 2.40E+06 & 6.69E+05 & 2.51E+05 & 27 & 6.79 & 2.12 & 0.88 & 8 & 31-May & 2:13 & 65 & -7 & M9.3 & 360 & 1835\\
44 & 2003 & 26-Oct & 17:40 & 1.81E+08 & 1.84E+07 & 1.75E+06 & 3.38E+05 & 466 & 42.6 & 3.78 & 0.8 & 11 & 26-Oct & 17:21 & 38 & 2 & X1.2 & 171 & 1537\\
45 & 2003 & 2-Nov & 17:20 & 1.43E+09 & 1.95E+08 & 3.22E+07 & 9.18E+06 & 1510 & 476 & 115 & 49.4 & 11 & 2-Nov & 17:03 & 56 & -14 & X8.3 & 360 & 2598\\
46 & 2003 & 4-Nov & 21:40 & 2.14E+08 & 3.26E+07 & 3.80E+06 & 7.76E+05 & 353 & 59.3 & 6.85 & 1.33 & 11 & 4-Nov & 19:29 & 83 & -19 & X28 & 360 & 2657\\
47 & 2004 & 19-Sep & 17:25 & 2.03E+07 & 3.02E+06 & 3.91E+05 & 9.35E+04 & 57.3 & 8.87 & 1.49 & 0.37 & 11 & 19-Sep & 16:46 & 58 & 3 & M1.9 & ..... & .....\\
48 & 2004 & 10-Nov & 3:05 & 2.81E+08 & 3.38E+07 & 4.24E+06 & 1.07E+06 & 424 & 49.4 & 7.52 & 2.42 & 11 & 10-Nov & 1:59 & 49 & 9 & X2.5 & 360 & 3387 \\
49 & 2005 & 17-Jan & 12:25 & 2.44E+09 & 6.04E+08 & 7.84E+07 & 1.32E+07 & 5040 & 1330 & 166 & 28.1 & 11 & 17-Jan & 6:59 & 25 & 15 & X3.8 & 360 & 2547 \\
50 & 2005 & 20-Jan & 6:40 & 8.13E+08 & 3.83E+08 & 1.73E+08 & 8.13E+07 & 1860 & 1550 & 968 & 652 & 11 & 20-Jan & 6:36 & 61 & 14 & X7.1 & 360 & 3256 \\
51 & 2005 & 22-Aug & 19:10 & 2.92E+08 & 1.65E+07 & 1.22E+06 & 2.56E+05 & 337 & 27.2 & 2.06 & 0.37 & 11 & 22-Aug & 16:46 & 65 & -13 & M5.6 & 360 & 2378\\
52 & 2006 & 13-Dec & 2:35 & 4.62E+08 & 1.67E+08 & 5.96E+07 & 2.36E+07 & 698 & 372 & 187 & 88.7 & 11 & 13-Dec & 2:14 & 23 & -6 & X3.4 & 360 & 1774\\
53 & 2006 & 14-Dec & 22:40 & 3.03E+07 & 7.32E+06 & 1.76E+06 & 5.57E+05 & 215 & 42.3 & 8.07 & 2.38 & 11 & 14-Dec & 21:07 & 46 & -6 & X1.5 & 360 & 1042\\
54 & 2011 & 7-Jun & 6:55 & 4.93E+07 & 1.73E+07 & 5.74E+06 & 2.22E+06 & 72.87 & 25.84 & 10.81 & 4.53 & 13 & 7-Jun & 6:16 & 54 & -21 & M2.5 & 360 & 1255 \\
55 & 2011 & 4-Aug & 4:05 & 1.09E+08 & 1.35E+07 & 2.30E+06 & 6.56E+05 & 80.05 & 22.04 & 5.48 & 1.8 & 13 & 4-Aug & 3:41 & 36 & 19 & M9.3 & 360 & 1315 \\
56 & 2011 & 9-Aug & 8:10 & 1.11E+07 & 3.56E+06 & 1.11E+06 & 4.34E+05 & 26.92 & 15.83 & 6.52 & 2.67 & 13 & 9-Aug & 7:48 & 69 & 17 & X6.9 & 360 & 1610\\
57 & 2012 & 23-Jan & 4:10 & 4.81E+09 & 4.41E+08 & 1.76E+07 & 1.50E+06 & 3895.4 & 447.84 & 20.63 & 2.39 & 13 & 23-Jan & 3:38 & 25 & 18 & M8.7 & 360 & 2175\\
58 & 2012 & 27-Jan & 17:55 & 8.33E+08 & 1.43E+08 & 2.20E+07 & 5.85E+06 & 795.5 & 126.43 & 29.96 & 11.88 & 13 & 27-Jan & 17:37 & 71 & 27 & X1.7 & 360 & 2508 \\
59 & 2012 & 13-Mar & 17:35 & 1.68E+08 & 1.64E+07 & 2.40E+06 & 6.66E+05 & 468.77 & 64.49 & 8.91 & 1.89 & 13 & 13-Mar & 17:12 & 59 & 19 & M7.9 & 360 & 1884\\
60 & 2012 & 17-May & 1:30 & 1.02E+08 & 3.03E+07 & 1.05E+07 & 4.24E+06 & 255.44 & 123.65 & 54.54 & 20.44 & 13 & 17-May & 1:25 & 76 & 11 & M5.1 & 360 & 1582\\
61 & 2012 & 6-Jul & 23:55 & 1.67E+07 & 2.56E+06 & 5.05E+05 & 1.63E+05 & 25.4 & 5.49 & 1.13 & 0.37 & 13 & 6-Jul & 23:01 & 51 & -17 & X1 & 360 & 1828\\
62 & 2013 & 22-May & 14:20 & 6.35E+08 & 7.97E+07 & 9.34E+06 & 2.18E+06 & 1196.6 & 121.05 & 12.73 & 3.4 & 15 & 22-May & 13:08 & 70 & 15 & M5.0 & 360 & 1466\\
63 & 2014 & 20-Feb & 8:15 & 3.78E+06 & 7.70E+05 & 2.10E+05 & 8.10E+04 & 22.25 & 7.79 & 2.24 & 0.69 & 13 & 20-Feb & 7:26 & 73 & 15 & M3.0 & 360 & 948\\
64 & 2014 & 18-Apr & 13:40 & 6.58E+07 & 4.86E+06 & 7.77E+05 & 2.37E+05 & 58.47 & 6.85 & 1.55 & 0.66 & 13 & 18-Apr & 12:31 & 34 & -2 & M7.3 & 360 & 1203\\
65 & 2017 & 06-Sep & 12:35 & 2.98E+08 & 1.11E+07 & 8.59E+05 & 2.28E+05 & 41.38 & 4.95 & 1.4 & 0.62 & 13 & 06-Sep & 11:53 & 33 & -8 & X9.3 & 360 & 1571\\ \hline
\end{tabular}%
}
\end{table*}

\end{document}